\documentclass[aps,pra,twocolumn,groupedaddress,floatfix]{revtex4-1}

\usepackage{graphicx}
\usepackage{dcolumn}
\usepackage{bm}
\usepackage{amsmath}
\usepackage{xcolor}
\usepackage{braket}

\usepackage{amsfonts}
\usepackage{blkarray}
\usepackage{amssymb}
\usepackage{multirow}
\usepackage{mathrsfs}

\DeclareMathOperator{\Tr}{Tr}

\begin{document}

\title{Interatomic interaction effects on second-order  momentum correlations and Hong-Ou-Mandel 
interference of double-well-trapped ultra cold fermionic atoms}

\author{Benedikt B. Brandt}
\email{benbra@gatech.edu}
\author{Constantine Yannouleas}
\email{Constantine.Yannouleas@physics.gatech.edu}
\author{Uzi Landman}
\email{Uzi.Landman@physics.gatech.edu}

\affiliation{School of Physics, Georgia Institute of Technology,
             Atlanta, Georgia 30332-0430}

\date{2 January 2018}

\begin{abstract}
Identification and understanding of the evolution of interference patterns in two-particle momentum
correlations as a function of the strength of interatomic interactions are important in explorations of the 
nature of quantum states of trapped particles. Together with the analysis of two-particle spatial 
correlations, they offer the prospect of uncovering fundamental symmetries and structure of correlated 
many-body states, as well as opening vistas into potential control and utilization of correlated quantum 
states as quantum information resources. With the use of the second-order density matrix constructed via 
exact diagonalization of the microscopic Hamiltonian, and an analytic 
Hubbard-type model, we explore here the systematic evolution of characteristic interference patterns in the
two-body momentum and spatial correlation maps of two entangled ultracold fermionic atoms in a double well,
for the entire attractive- and repulsive-interaction range. We uncover statistics-governed bunching and 
antibunching, as well as interaction-dependent interference patterns, in the ground and excited states,
and interpret our results in light of the Hong-Ou-Mandel interference physics, 
widely exploited in photon indistinguishability testing and quantum information science.
\end{abstract}

\maketitle

\section{Introduction}

The rapid experimental progress in the field of ultracold atoms is enabling measurements with unprecedented
precision of fundamental many-body quantities such as higher-order correlations 
\cite{bloc05,bloc06.2,kauf14,bouc16,hodg17,schm17,berg17}, especially higher-order {\it momentum} 
correlations for interacting \cite{bouc16,hodg17,berg17} ultracold atoms in linear traps. 
The study of these correlations, with the full ability of tuning the interparticle interactions (utilizing
the Feshbach resonance technique) and under pristine environmental conditions, promises to deepen 
our understanding and potential technological control of quantum information processes \cite{bloc06} and 
physical phenomena, such as entanglement \cite{isla15} and generation of exotic many-body regimes 
(e.g., Tonks-Girardeau states \cite{bloc04}). However, in spite of the recent burgeoning experimental 
activities aiming at measuring higher-order momentum correlations \cite{bouc16,hodg17,schm17,berg17}, 
corresponding theoretical investigations are still lacking in many respects, apart from a couple of studies
\cite{bouc16,yann17}. 

In this paper, we study the systematic evolution of the properties and interference patterns of 2nd-order 
(two-particle) momentum correlations of two interacting 
(both distinguishable and indistinguishable) ultracold 
fermions in a double-well optical trap. To provide a complete picture, we go beyond the case of the ground 
singlet and 1st-excited triplet states and investigate in addition the cases of the 2nd and 3rd excited
states, both singlets. (This quartet of states can be mapped to a two-site Hubbard model; see below.)

Elucidating the 2nd-order momentum correlations associated with double-well trapping of two ultracold atoms
(without \cite{schm17} or with \cite{berg17} interactions) is currently attracting pioneering experimental 
interest, both planned \cite{schm17} and preparatorily achieved \cite{berg17}. These 
experimental efforts are motivated by the unprecedented tunability of: (i) the confining external optical 
potential and the dynamical imprinting of a relative phase difference between the two wells \cite{schm17},
and (ii) the two-body contact interaction via a combination of Feshbach and confinement-induced resonances
\cite{joch12,joch15}.

The double-well two-particle unit \cite{joch15,yann15} is expected to be a central component for building 
more complex quantum-computer and quantum-information architectures, and detailed knowledge of the 
associated 2nd-order momentum correlations is emerging as an indispensable tool towards implementation of 
these endeavors \cite{schm17,berg17}. In this context, recent work \cite{schm17,kauf14} investigates the 
double-well atomic dimers treating them as purely photonic analogs (i.e., omitting or minimizing the role 
of interparticle interaction). The interparticle interaction, however, is an essential factor in particle 
assemblies and the desirability of a full understanding of its effects can hardly be overestimated. 
\textcolor{black}{
The seminal optical Hong-Ou-Mandel (HOM) second-order-interference experiment \cite{hong87,ou07}, 
widely exploited in photon indistinguishability testing and quantum information science, 
spawned extensions of such interference phenomena to 
electrons \cite{liu98,bocq13} and bosonic atoms \cite{kauf14,aspe15}. Here we further interpret our 
correlations results for ultracold fermions in light of the HOM physics.}

The much sought-after deeper understanding of the double-well fermionic dimer is achieved below through 
employment of an exact configuration-interaction (CI) method for solving the two-body problem, in 
conjunction with a modified Hubbard-type analytic modeling that allows a synoptic interpretation of the 
properties and interference patterns of the microscopic, numerically CI-derived, two-particle momentum 
correlations. 

\begin{figure*}[t]
\includegraphics[width=15.5cm]{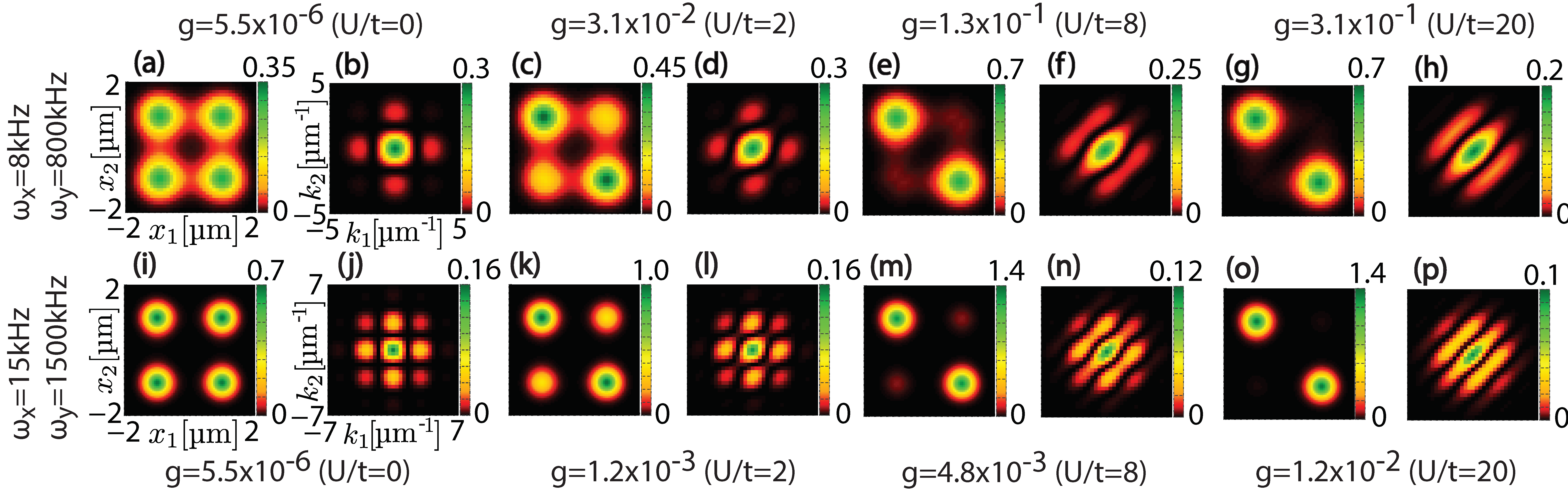} 
\caption{Ground-state CI-calculated spatial and momentum correlation maps for two fermions in a double 
well, as a function of the two-body interaction strength $g$. The interwell distance is $d=2$ $\mu$m.
The results in the upper and lower rows correspond to two different choices of the confining harmonic  
frequencies along the inter-well direction ($x$) and in the transverse one ($y$); for both cases  
$\omega_x/\omega_y = 1/100$. Because of the quasilinear nature of the system, here and for all 2D CI-derived
correlations, the maps are drawn for $y_1=y_2=0$ for the spatial correlations and for $k_1^y=k_2^y=0$ for 
the momemtum correlations. Note that we drop for convenience the superscript $x$ and use $k_i=k_i^x$, where
$i=1,2$ denotes the index numbering the two particles. This yields the plotted correlation maps for
the position $(x_1, x_2)$ and momentum $(k_1, k_2)$ variables along the $x$-direction connecting
the two wells.}
\label{fig1}
\end{figure*}

\section{Theory essentials}

To implement the microscopic CI method, we start by considering the two-dimensional (2D) Hamiltonian of two 
interacting ultracold fermions,
\begin{align}
H_{\rm MB}= H(1)+ H(2) + V({\bf r}_1,{\bf r}_2),
\label{mbh}
\end{align}

\noindent where $H(i)$ represents the single particle part of the many-body Hamiltonian and 
$V({\bf r}_1,{\bf r}_2)$ represents the interaction term, with ${\bf r}_i \equiv (x_i,y_i)$, 
$i=1,2$, being the space coordinates of the first and second particle. 
The single particle part $H(i)$ of the Hamiltonian contains the kinetic energy term and a single-particle 
external confining potential; in this paper we consider a double-well confinement.

The double-well external confining potential has been extensively described in Refs.\ \cite{yann15,yann16}. 
The relevant potential parameters are the inter-well spacing $d$ along the $x$-direction, and
the value of $\epsilon_b$ (determining the interwell barrier height) which is taken to be $0.5$ 
throughout the paper. Each of the parabolic confining wells is characterized by two harmonic frequencies, 
$\hbar\omega_x$ (along the $x$-axis of the well) and $\hbar\omega_y$ (along the $y$ direction), resulting 
in a (quasi-onedimensional) needle-like shape confinement when $\hbar\omega_x << \hbar\omega_y$. In our 
calculations here, we consider two different sets of values, i.e., $\hbar\omega_x = 8$ kHz, 
$\hbar\omega_y = 800$ kHz and $\hbar\omega_x = 15$ kHz, $\hbar\omega_y = 1500$ kHz, both having the same
aspect ratio $\omega_x/\omega_y$; hereafter we drop for convenience the subscript $x$ and use
$\omega=\omega_x$). 

The short-range interatomic interaction term is given by 
\begin{align}
V({\bf r}_1,{\bf r}_2)=\frac{g}{\sigma^2 \pi}e^{-({\bf r}_1-{\bf r}_2)^2/\sigma^2}.
\label{tbi}
\end{align}
\noindent In this paper we use $\sigma=0.01$ $\mu$m, yielding a ratio $\sigma/l_0 \sim 0.03$ for the case 
of $\hbar\omega=8$ kHz and $\sigma/l_0 \sim 0.022$ for the case of $\hbar\omega=15$ kHz; $l_0$ is the
oscillator length $l_0^2=\hbar/(M_{^6{\rm Li}}\omega)$, with $M_{^6{\rm Li}}=10964.90m_e$ being the mass 
of $^6{\rm Li}$; a pair of states out of the three lowest $^6{\rm Li}$ hyperfine states corresponds to two 
different spin states \cite{joch12}. The factors $\sigma/l_0$ are motivated by the need to model 
short-range, contact-type interactions. Any Gaussian width $\sigma$ that is sufficiently smaller than the 
harmonic oscillator length $l_0$ along the $x$-direction is suitable and yields essentially identical final 
results. Here we consider both mutually repelling ($g>0$) and attractive ($g<0$) particles and the tunable 
values of the interaction strength $g$ will be given in units of $\hbar \omega l_0^2$.

Because for $N=2$ fermions the spin variables separate from the space variables \cite{shul56}, the CI wave 
function has the product form $\Phi_{\rm CI}^{S,S_z}({\bf r}_1,{\bf r}_2) \chi(S,S_z)$, where $S$ and 
$S_z$ denote the total spin and its projection. As a result, for $N=2$ the spin-resolved and 
spin-unresolved two-body correlations are the same \cite{yann17} apart from an overall factor. Then the 
two-body space correlation is defined by
\cite{yann17}
\begin{align}
{\cal P}_{\rm CI}^{S,S_z}({\bf r}_1,{\bf r}_1',{\bf r}_2,{\bf r}_2')=
\Phi_{\rm CI}^{S,S_z\dagger}({\bf r}_1,{\bf r}_1')
\Phi_{\rm CI}^{S,S_z}({\bf r}_2,{\bf r}_2'),
\label{tbsc}
\end{align} 
\noindent while the two-body momentum correlation is given by the Fourier transform
\begin{align}
\begin{split}
{\cal G}_{\rm CI}^{S,S_z}({\bf k}_1,{\bf k}_2)= & \frac{1}{4\pi^2}
\int_{-\infty}^\infty e^{-i {\bf k}_1 \cdot ({\bf r}_1-{\bf r}_1')}
\int_{-\infty}^\infty e^{-i {\bf k}_2 \cdot ({\bf r}_2-{\bf r}_2')} \\
& \times
{\cal P}_{\rm CI}^{S,S_z}({\bf r}_1,{\bf r}_1',{\bf r}_2,{\bf r}_2')
d{\bf r}_1 d{\bf r}_1' d{\bf r}_2 d{\bf r}_2'.
\end{split}
\label{tbmc}
\end{align} 

\section{Analysis of the ground state}

In Fig.\ \ref{fig1} we plot the CI two-body correlations for two repelling fermions in their singlet
($S=0,S_z=0$) ground state as a function of the interaction strength $g$ (in units of 
$\hbar \omega l_0^2$); the interwell separation is $d=2$ $\mu$m. The values of $g$ are also expressed as 
the ratio $U/t$ between the on-site repulsion ($U$) and the intersite hopping parameter ($t$)
associated with the two-site Hubbard model (whose parameters have been extracted from the microscopic CI 
calculation; see Appendix \ref{a2}). Two different confining
harmonic potentials have been considered with energy spacings $\hbar \omega = 8$ kHz (top row) and
$\hbar \omega = 15$ kHz (bottom row). In all cases in this work, we show two-particle spatial 
correlation maps for $y_1=y_2=0$ and two-particle momentum correlation maps for $k^y_1=k^y_2=0$; we 
verified that similar results are obtained for other $y_1=y_2={\rm const.}$ and $k^y_1=k^y_2={\rm const.}$ 
values. Note that we drop for convenience the superscript $x$ and use $k_i=k_i^x$, where $i=1,2$ denotes 
the index numbering the two particles.

The spatial correlations for the above-noted two confining-potential energy spacings  
\ref{fig1}(a,c,e,g) and Figs.\ \ref{fig1}(i,k,m,o), respectively], exhibit similar 
behavior as $g$ (or $U/t$) increases, transforming from a four-hump pattern in a square formation
to a two-hump one along the $x_1+x_2=0$ diagonal (referred to here as ``antidiagonal''). Naturally in the 
non-interacting limit [$g \sim 0$, $U/t \sim 0$, Fig.\ \ref{fig1}(a) and Fig.\ \ref{fig1}(i)], the two 
humps located along the $x_1-x_2=0$ diagonal (referred to here as ``main diagonal'') are due to the double 
occupancy (involving both the $\uparrow$ and $\downarrow$ spins) of the 
lowest symmetric single-particle orbital of the double well, which in the Hubbard modeling translates 
into double occupancy of each site. As $g$ increases, the double-occupancy humps along the main
diagonal progressively shrink, and they eventually vanish in the strong-repulsion regime [see case for 
$U/t=20$ in Fig.\ \ref{fig1}(g) and Fig.\ \ref{fig1}(o)].

\begin{figure}[b]
\includegraphics[width=6.cm]{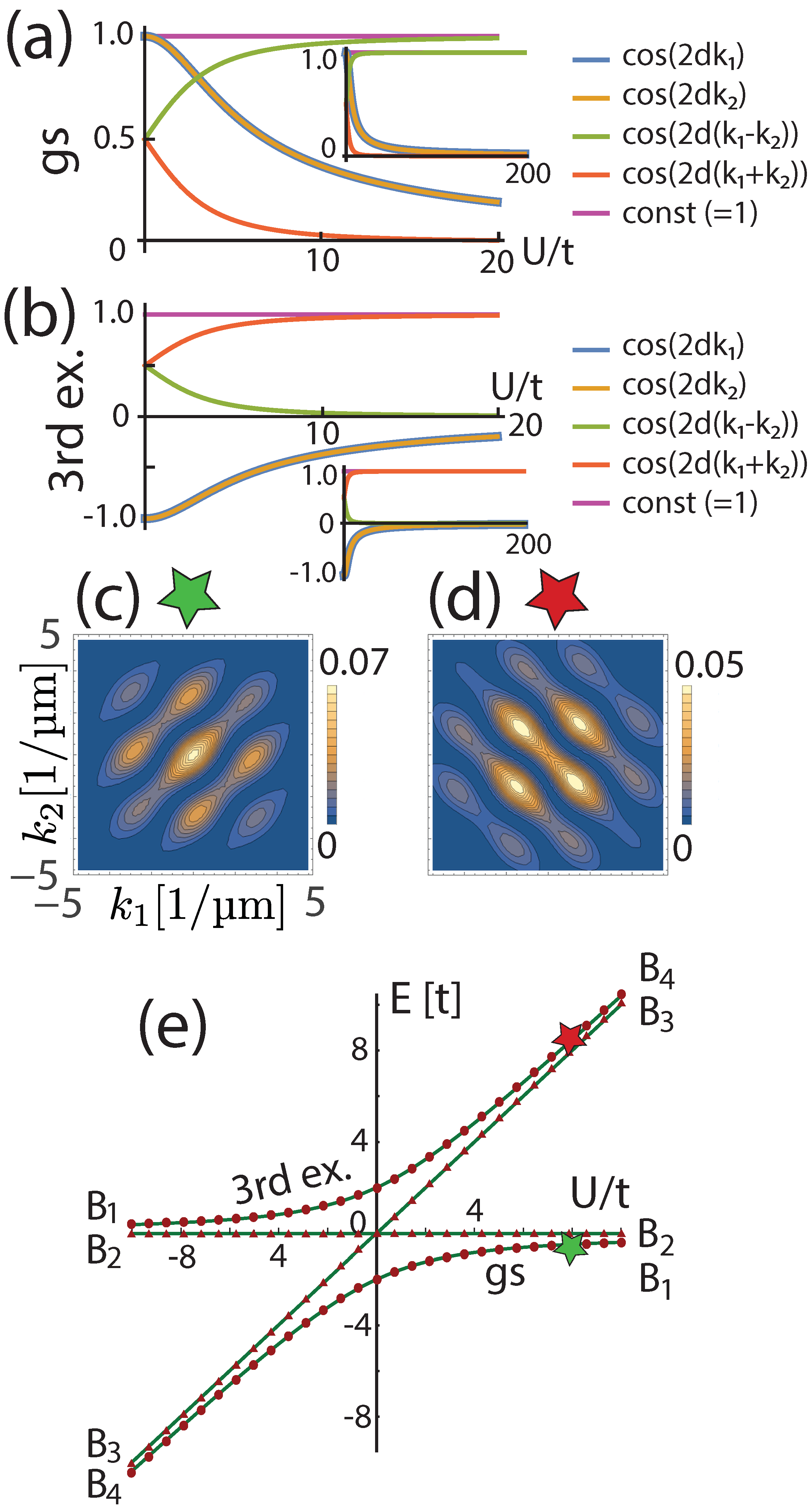}
\caption{Plots of analytic weights of the ground (a) and 3rd excited (b) states [both singlets, see Eqs.\ 
(\ref{hubgs}) and (\ref{hub3ex})] of the various contributing terms in the two-body momentum correlations 
as a function of the strength of the Hubbard interaction parameter $U/t$. The contributions to the various 
terms in Eqs.\ (\ref{hubgs}) and (\ref{hub3ex}) are identified by different colors as indicated on the 
right in (a) and (b). (c-d) The Hubbard momentum maps at $U/t=8$ for the ground (green star) and 3rd 
excited (red star) states. (e) The energy spectrum (solid lines) of the two-site Hubbard model covering
both the attrative ($U/t <0$) and repulsive ($U/t >0$) ranges. The symbols $B_j$, $j=1,\cdots,4$ denote the
four Bell states at $U/t \rightarrow \pm \infty$. The red dots are the corresponding 
microscopic CI energies. The Hubbard model in (c-e) corresponds to the CI calculation with $d=2$ $\mu$m and 
$\hbar\omega=15$ kHz. Hubbard-model analytic two-particle spatial and momentum correlation maps for the 
ground state and the three lowest excited states for the repulsive ($U/t = 8$) and attractive ($U/t =-8$) 
cases are shown in Figs.\ \ref{figs2}, \ref{figs3}, and \ref{figs4} in Appendix \ref{a3}. 
Note the reversal of the energy-ordering of the  
Bell states for the Hubbard $U/t \rightarrow -\infty$ and $U/t \rightarrow +\infty$ limits. 
}
\label{fig2}
\end{figure}

The evolution of the two-body momentum correlations [Figs.\ \ref{fig1}(b,d,f,h) and Figs.\ 
\ref{fig1}(j,l,n,p)] is more complex. At the non-interacting limit [Fig.\ \ref{fig1}(b) and Fig.\ 
\ref{fig1}(j)], a plaid pattern of circular humps is evident. As a function of increasing $g$, the plaid
pattern distorts and transforms into interference fringes exhibiting elongated maxima along and parallel to
the main diagonal ($k_1-k_2=0$); the associated valleys (minima) of this pattern appear along the 
antidiagonal ($k_1+k_2=0$). This interference pattern is well developed for $U/t=20$ for which
the residues of the $U=0$ circular humps only minimally distort the parallel fringes. We checked that
the $U=0$ circular humps do not survive for larger values of $g$ (or $U/t$). 

Furthermore, there is a prominent qualitative difference between the top- ($\hbar\omega=8$ kHz 
confinement) and bottom-row ($\hbar\omega=15$ kHz confinement) momentum maps. Indeed for $\hbar\omega=15$ 
kHz (bottom row), there are more individual features (humps or fringes) compared to the case of 
$\hbar\omega=8$ kHz (top row). In particular, we note for the 
independent particle case that there are nine visible 
humps in Fig.\ \ref{fig1}(j) compared to four humps in Fig.\ \ref{fig1}(b), with the additional maxima
in Fig.\ \ref{fig1}(j) revealing for $U=0$ enhanced correlations between particles with equal momenta, 
regardless of their signs. Similarly for $U/t = 20$ we find five visible fringes 
in Fig.\ \ref{fig1}(p) compared to three in Fig.\ \ref{fig1}(h), with the added fringes in Fig.\ 
\ref{fig1}(p) revealing correlations between particles having the same, but of opposite sign, momenta. 

To gain insights about the systematics in the evolution of the momentum maps, we model the fermion
single-particle space orbitals as displaced Gaussian functions centered at each well. Taking account of the
spin, the ensuing Gaussian-type spin-orbitals are used to form Slater determinants according to the spin 
eigenfunctions of the corresponding two-site Hubbard model (with parameters $U$ and $t$ extracted from the 
CI calculations; see Appendix \ref{a2}). This procedure endows the Hubbard model eigenvector solutions 
with the (otherwise absent) spatial degrees of freedom; see Appendix \ref{a1}. 
Considering the strictly one-dimensional case 
along the $x$-axis and applying the definition in Eq.\ (\ref{tbmc}) to these modified Hubbard-model 
solutions, one obtains for the two-body momentum correlation of the singlet ground state
\begin{align}
\begin{split}
& {\cal G}_{\rm Hub,gs}^{S=0,S_z=0}(k_1,k_2) \propto 
\frac{2 s ^2 e^{-2 s^2 (k_1^2+k_2^2)}}
{\pi (\mathcal{U} \mathcal{Q}(\mathcal{U})+16)} \\
& \times \bigg( (\mathcal{U}\mathcal{Q}(\mathcal{U})+8) 
\cos (2 d (k_1-k_2))+8 \cos (2 d (k_1+k_2)) \\ 
& +4\mathcal{Q}(\mathcal{U}) \cos (2 d k_1)+ 4\mathcal{Q}(\mathcal{U}) 
\cos (2 d k_2)+\mathcal{U}\mathcal{Q}(\mathcal{U}) +16 \bigg),
\end{split}
\label{hubgs}
\end{align}
\noindent 
where ${\cal U}=U/t$, $\mathcal{Q}(\mathcal{U})=\sqrt{\mathcal{U}^2+16}+\mathcal{U}$, $s$ is the width of
the Gaussian orbital, and $d$ is the interwell distance. Eq.\ (\ref{hubgs}) is valid for both 
negative (${\cal U} \leq 0$, attractive) and positive (${\cal U} > 0$, repulsive) values; similarly,
the expressions in Eqs.\ (\ref{hub1ex})-(\ref{hub3ex}) below are valid in the whole range 
$-\infty < {\cal U} < +\infty$. Note that ${\cal Q}(-{\cal U}) = {\cal P}({\cal U}) \equiv
\sqrt{ {\cal U}^2+16 }-{\cal U}$ and that ${\cal U}{\cal P}({\cal U}) \rightarrow 8$ when
${\cal U} \rightarrow \infty$.

In Eq.\ (\ref{hubgs}), four specific $\cos$ 
terms contribute, displaying oscillations along the main diagonal $(k_1-k_2)$, the antidiagonal
($k_1+k_2$), and the two axes ($k_1$ and $k_2$). These four terms are supplemented with a constant fifth,
circularly-symmetric contribution. Each of these terms is damped by an exponential prefactor
$e^{-2 s^2 (k_1^2+k_2^2)}$ whose range ($1/2s^2$) depends on the width $s$ of the displaced Gaussian 
orbitals. This fact accounts for the different number of visible individual features (circular humps or 
fringes) in the CI momentum maps between the top and bottom row of Fig.\ \ref{fig1}. Indeed a narrower 
confining potential (i.e., the one with $\hbar\omega= 15$ kHz) results in a smaller spatial extent of the 
associated single-particle states compared to a wider confining potential (i.e., the one with 
$\hbar\omega=8$ kHz); the oscillator length (and thus $s$) is inversely proportional to $\sqrt{\omega}$, 
leading to a damping range $1/2s^2 \propto \omega/2$.

The evolution of the analytic weights for the Hubbard ground-state [coefficients in front of the four 
$\cos$ terms plus the constant term in Eq.\ (\ref{hubgs}) without the overall common factor 
$2s^2e^{-2 s^2 (k_1^2+k_2^2)}/\pi$] are plotted as a function of $U/t$ in Fig.\ \ref{fig2}(a);
the spectra for the ground and three lowest excited states are displayed in Fig.\ \ref{fig2}(e). 
The variation of these weights provides a direct interpretation of
the evolution of the CI momemtum maps in Fig.\ \ref{fig1}. In fact for non-interacting fermions 
($g \sim 0$ or $U=0$), all five terms contribute in a substantial way in the sum of Eq.\ (\ref{hubgs}),
and this leads to the plaid pattern in Figs.\ \ref{fig1}(b) and \ref{fig1}(j). For strong $g$ (or high
$U/t$), only two contributions survive, i.e., the constant and the $\cos(2d(k_1-k_2))$ terms with equal 
weights. The corresponding Hubbard momentum map (for $U/t=8$) plotted in Fig.\ \ref{fig2}(c) 
[see lower, green star in Fig.\ \ref{fig2}(e)] is found to 
agree with the pattern and orientation of the fringes observed in the CI-calculated maps in Figs.\ 
\ref{fig1}(f) and \ref{fig1}(n). The analytic parameter $s$ in Fig.\ \ref{fig2}(c) was adjusted to 
correspond to a potential well with a steeper confinement (i.e., $\hbar\omega=15$ kHz); in this case there 
are five visible fringes in Fig.\ \ref{fig2}(c) precisely as in the CI case in Fig.\ \ref{fig1}(n). Note 
that in the strong-interaction case, the two-term $1+\cos(2d(k_1-k_2))=2\cos^2(d(k_1-k_2))$ pattern can be
reproduced also using \cite{yann17} a Heisenberg-Hamiltonian modeling.

\begin{figure}[t]
\includegraphics[width=6.5cm]{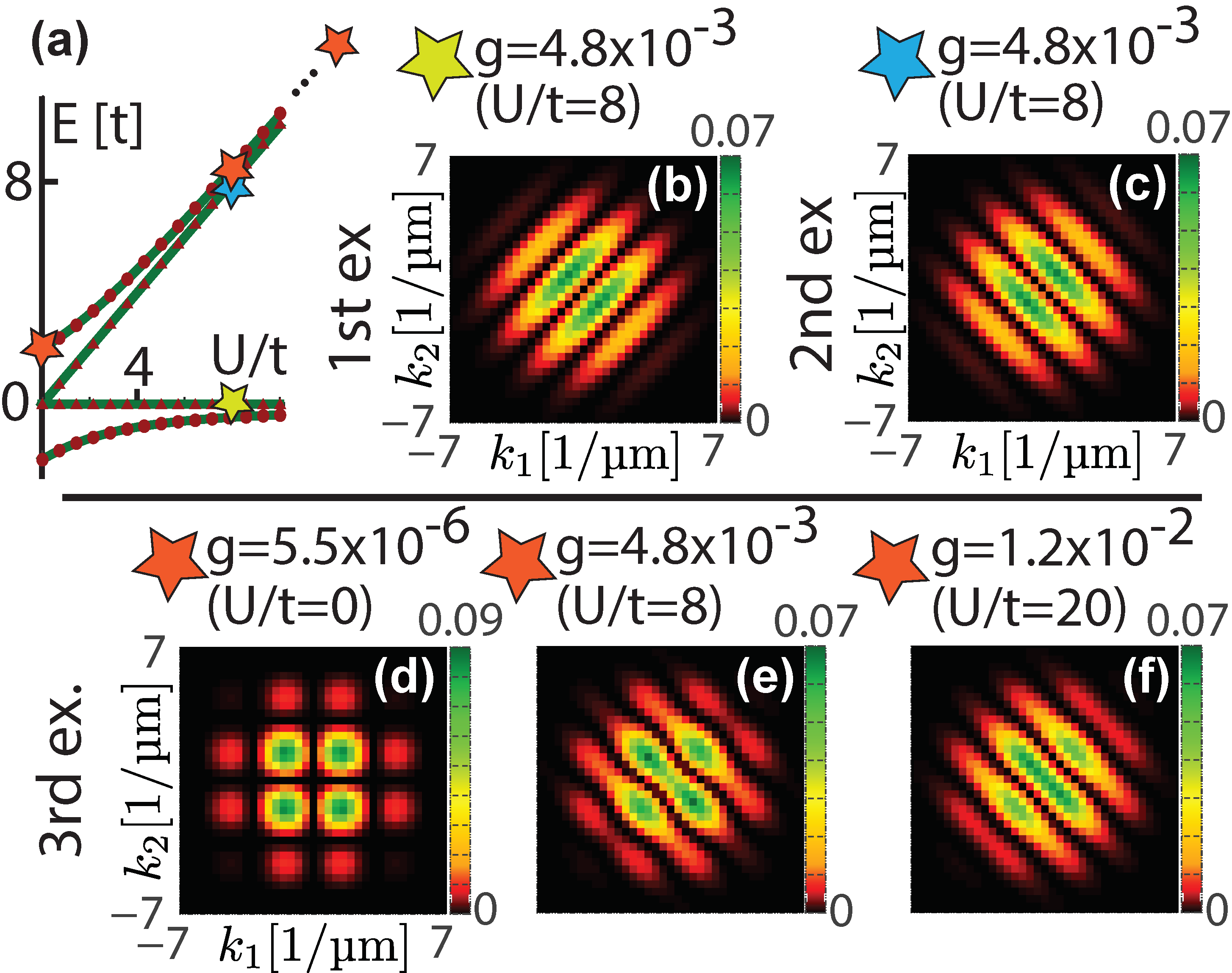}
\caption{CI momentum correlation maps for two fermions in a double well associated with the first three 
excited states, denoted by a yellow, blue, and orange star, respectively. The interwell distance is 
$d=2$ $\mu$m, and the steeper potential confinement ($\hbar \omega=15$ kHz) is used. The energy spectrum of
the corresponding two-site Hubbard model is plotted in (a). The stars in (a) indicate the specific
values of $U/t$ (corresponding to particular $g$'s) for which the CI momentum maps for the 1st excited (b),
2nd excited (c), and 3rd excited (d-f) states were calculated. $g$ is in units of $\hbar \omega l_0^2$.
The red dots or triangles in (a) are the corresponding microscopic CI energies.}
\label{fig3}
\end{figure}

\section{Analysis of excited states}

CI momentum maps for the first three excited states are displayed in Fig.\ \ref{fig3}. For the 1st
(triplet) and 2nd (singlet) excited states [see the the yellow and blue stars in Fig.\ \ref{fig3}(a)], 
the momentum correlation maps are independent of the interparticle interaction (because of the 
wave function nodal structure for these excited states), and thus we display for these states the results  
for a single value of the interaction  $(U/t=8)$; see Figs.\ \ref{fig3}(b) and \ref{fig3}(c), 
respectively. Indeed the analytic expressions of the corresponding two-site 
Hubbard model contain only a single sinusoidal term, independent of the parameter $U/t$, namely
\begin{align}
{\cal G}_{\rm Hub,1st~ex}^{S=1,S_z=0}(k_1,k_2) \propto   
\frac{4s^2 e^{-2s^2 (k_1^2+k_2^2)} \sin^2 (d (k_1-k_2))}{\pi },
\label{hub1ex}
\end{align}
\noindent and
\begin{align}
{\cal G}_{\rm Hub,2nd~ex}^{S=0,S_z=0}(k_1,k_2) \propto
\frac{4s^2 e^{-2s^2 (k_1^2+k_2^2)} \sin^2 (d (k_1+k_2))}{\pi }.
\label{hub2ex}
\end{align}
\noindent We have checked that Eq.\ (\ref{hub1ex}) applies to the other two $S=1$, $S_z=\pm 1$ triplet
states as well.

In Fig.\ \ref{fig3}(b) (1st CI excited state), the valley of vanishing values lies along the main diagonal 
(antibunching behavior), a fact that reflects the Pauli exchange principle which comes into play for a 
triplet state ($S=1$, antisymmetric space wave function). We further note that in Fig.\ \ref{fig3}(c) (2nd 
CI excited state), the orientation of the fringes is perpendicular to that in Fig.\ \ref{fig3}(b), a 
behavior that reflects the $\sin^2 (d (k_1+k_2))$ oscillatory pattern in Eq.\ (\ref{hub2ex}) (associated 
with the $B_3$, $S=0$ symmetric in space Bell state) versus the $\sin^2 (d (k_1-k_2))$ one in Eq.\ 
(\ref{hub1ex}).  

Figs.\ \ref{fig3}(d-f) describe the evolution with increasing repulsion of the CI momentum maps for the 3rd
excited state [orange stars on the upper curve in Fig.\ \ref{fig3}(a)]. This evolution can be interpreted 
by considering the corresponding analytic two-site Hubbard momentum correlation 
\begin{align}
\begin{split}
& {\cal G}_{\rm Hub,3rd~ex}^{S=0,S_z=0}(k_1,k_2) \propto
\frac{2 s ^2 e^{-2 s^2 (k_1^2+k_2^2)}}
{\pi (16-\mathcal{U} \mathcal{P}(\mathcal{U}))} \\
& \times \bigg( (8-\mathcal{U}\mathcal{P}(\mathcal{U}))
\cos (2 d (k_1-k_2)) + 8 \cos (2 d (k_1+k_2)) \\
& -4\mathcal{P}(\mathcal{U}) \cos (2 d k_1)- 4\mathcal{P}(\mathcal{U})
\cos (2 d k_2)+ 16-\mathcal{U}\mathcal{P}(\mathcal{U}) \bigg),
\end{split}
\label{hub3ex}
\end{align}
where as aforementioned $\mathcal{P}(\mathcal{U})=\sqrt{\mathcal{U}^2+16}-\mathcal{U}$.

The analytic weights of the five contributing terms in Eq.\ (\ref{hub3ex}) as a function of $U/t$ are 
plotted in Fig.\ \ref{fig2}(b). \big(As aforementioned ${\cal U}{\cal P}({\cal U}) \rightarrow 8$ when 
${\cal U} \rightarrow \infty$.\big) For the non-interacting limit ($U=0$), all five terms contribute and
yield a plaid pattern [see Fig.\ \ref{fig3}(d)], as was also the case for the singlet ground state. For 
very strong interactions only the two contributions $1+\cos(2d(k_1+k_2))=2\cos^2(d(k_1+k_2))$ survive; see 
Fig.\ \ref{fig3}(f) corresponding to $U/t=20$. For an intermediate $U/t=8$, Eq.\ (\ref{hub3ex})
is plotted in Fig.\ \ref{fig2}(d) [see upper, red star in Fig.\ \ref{fig2}(e)], 
exhibiting fringes with a dominant $1+\cos(2d(k_1+k_2))$ behavior, which
is however distorted by residual humps due to the other three weaker terms. The Hubbard pattern in Fig.\ 
\ref{fig2}(d) agrees very well with the CI momentum map in Fig.\ \ref{fig3}(e); for additional two-particle
spatial and momentum correlation maps according to the Hubbard model, see Figs.\ \ref{figs2}, \ref{figs3}, 
and \ref{figs4} in Appendix \ref{a3}. 

\textcolor{black}{
\section{Entanglement aspects and connection to the Hong-Ou-Mandel interference physics}
The Hubbard-model eigenstates (see details in Appendix \ref{a41}), 
are a superposition of the four maximally 
entangled Bell states $B_1=(|LR\rangle-|RL\rangle)/\sqrt{2}$, $B_2=(|LR\rangle+|RL\rangle)/\sqrt{2}$,
$B_3=(|LL\rangle-|RR\rangle)/\sqrt{2}$, and $B_4=(|LL\rangle+|RR\rangle)/\sqrt{2}$, where $|L\rangle$,
$|R\rangle$ are, repectively, the single-particle states (including spin) in the left or right well; the 
superposition coefficients depending on the parameter ${\cal U}$. This is illustrated in Fig.\ 
\ref{fig2}(e), where the corresponding Bell states at ${\cal U} \rightarrow \pm \infty$ are explicitly
denoted. The first and second excited states are the pure Bell states $B_2$ and $B_3$,
respectively, for any ${\cal U}$. The Hong-Ou-Mandel \cite{hong87} interference phenomena are 
related to the coincidence probability $P_{11}$ of having two particles in the $B_1$ (indistinguishable 
bosons \cite{kauf14,aspe15}, $P_{11}=0$) or $B_2$ state (indistinguishable fermions \cite{liu98,bocq13}, 
$P_{11}=1$ due to the Pauli exclusion principle).} 

\begin{figure}[t]
\includegraphics[width=6.5cm ]{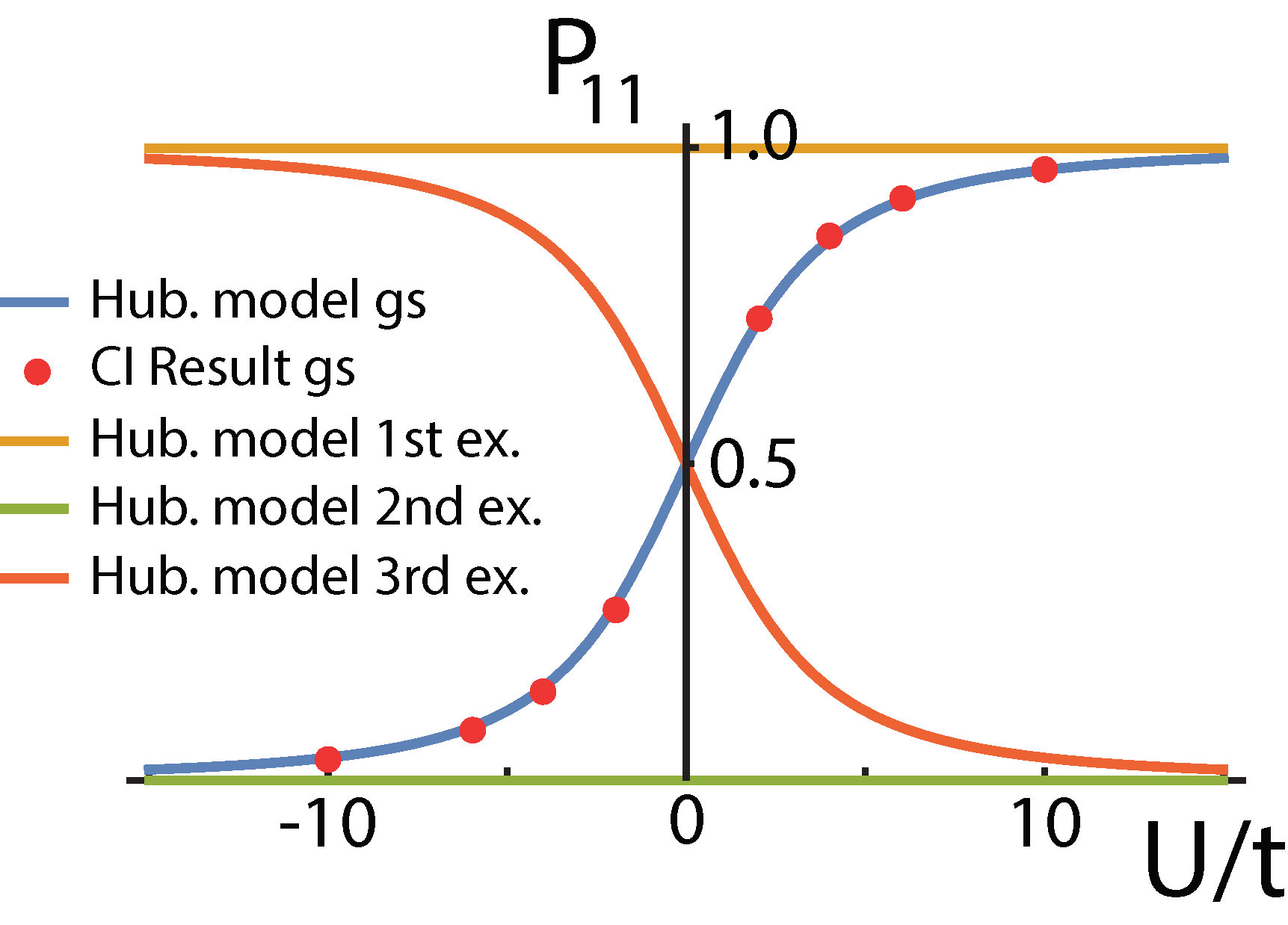} 
\caption{$P_{11}$ as a function of ${\cal U}=U/t$. $d=2$ $\mu$m. 
The red dots are corresponding CI results.}
\label{fig4}
\end{figure}

\textcolor{black}{
In our treatment, 
$P_{11}$ can be related to the second-order spatial and momentum correlations through the diagonal elements
of the two-particle density matrix $\rho_{ijkl}$ which decomposes the second-order correlation maps to 
left-right ($L$,$R$) components. From the momentum correlation maps, and using the Hubbard modeling for 
simplicity, one has 
\begin{align}
\begin{split}
& \mathcal{G}_{\rm Hub}(k_1,k_2) = \sum_{i,j,k,l=L,R}\eta_{ijkl}^{\rm Hub,2nd}(k_1,k_2)=\\
& \sum_{i,j,k,l=L,R}\rho_{ijkl}^{\rm Hub,2nd}\psi_{i\uparrow\
}(k_1)\psi_{j\downarrow}(k_2)\psi_{k\uparrow}^\dagger(k_1)\psi_{l\downarrow}^\dagger(k_2).
\end{split}
\label{deco}
\end{align}     
\noindent 
The explicit expressions for $\rho_{ijkl}$ for the four Hubbard states are given in Appendices
\ref{a43}, \ref{a44}, \ref{a45}, and \ref{a46}. $P_{11}=\rho_{LRLR}+\rho_{RLRL}$; Fig.\ \ref{fig4} 
displays the dependence of $P_{11}$ on ${\cal U}$.
}

\textcolor{black}{
Additional HOM [and also Handbury Brown-Twiss \cite{bloc05,bloc06.2,aspe07,ou88} (HBT)] 
aspects can be evoked based on the role 
played by the four Bell states in our approach. Developing corresponding experimental protocols that will 
test, among other possibilities, the interplay of beam splitters and interaction effects is beyond the   
scope of this paper. However, we mention here two possible paths. The first is the measurement of
spatial noise \cite{altm04} in the particle counts in the image of the expanding cloud of the two ultracold
atoms; this image reflects in space the momentum correlation maps. Such measurements along the main 
diagonal or antidiagonal of the image will correspond to the observation of both HOM antibunching and 
bunching types with fermions when using the first or second excited states, respectively; see
Figs.\ \ref{fig3}(b) and \ref{fig3}(c). This will follow 
the spirit of Refs.\ \cite{liu98,bocq13} that address the fermionic case 
for electrons by measuring current noise in mesoscopic semiconductors \cite{blan00}. Away from the two 
diagonals, the noise measurements may be associated with oscillatory HBT interference reflecting
the distance $d$ between the two wells \cite{bloc05,bloc06.2,aspe07,ou88}. 
Furthermore, if the left- or right-well 
provenance of the particles can be determined, noise measurements associated with
the components $\eta_{ijkl}$ of the momentum correlation maps [see Eq.\ (\ref{deco})], 
could be performed, yielding additional pathways for exploration of particle interference effects.
The second path relates to entanglement aspects
by using the density matrix $\rho_{ijkl}$ in the spirit of Refs.\ \cite{isla15,schm17}.
}

\section{Summary}

In conclusion, with the use of two-particle density matrix constructed via 
configuration-interaction exact diagonalization of the microscopic Hamiltonian, we have explored here the 
systematic evolution of characteristic, damped, interference patterns in the two-particle momentum and 
spatial correlation maps of two ultracold fermionic atoms trapped in a double-well potential, over the 
entire range of variation of the contact (both repulsive and attractive) 
interatomic interaction strength. For the singlet 
ground state the two-body momentum maps were found to transform from a square-plaid pattern [Figs.\ 
\ref{fig1}(b) and \ref{fig1}(j)] for vanishing interparticle interaction, to a system of striped 
interference fringes oriented in the direction parallel to the main diagonal of the square two-particle 
map [Figs.\ \ref{fig1}(h) and \ref{fig1}(p)]. The most intense fringe lies along the main diagonal
indicating bunching. Our theoretical results (Fig.\ \ref{fig1}, top row) agree well with the 
evolution (found with increasing strength) of preparatory experimentally measured \cite{berg17} momentum 
correlation maps \cite{note1}. We have  also analyzed two-body momentum correlation maps for low-lying 
excited states (Figs.\ \ref{fig2} and \ref{fig3}). The triplet excited state is associated with 
antibunching [see Fig.\ \ref{fig3}(b)]. A derived modified-Hubbard-type effective model, 
incorporating spatial degrees of freedom (i.e., interwell distance and particle localization length), in
addition to the customary on-site $U$ and hopping ($t$) Hubbard-Hamiltonian parameters 
(determined in each case from the CI results), has been found to reproduce well the microscopic CI results.
Importantly, this development allowed us to uncover analytic expressions capturing the full evolution of 
the two-particle momentum correlation maps over the entire range of interparticle interactions -- from the 
non-interacting regime ($U/t = 0$, with  substantial ground-state site-double-occupancy contributions), to 
the Mott insulating regime with large $U/t$.\\ 
~~~\\

\begin{acknowledgments}
This work has been supported by a grant from the Air Force Office of Scientific Research (USA) under 
Award No. FA9550-15-1-0519. Calculations were carried out at the GATECH Center for Computational Materials 
Science.
\end{acknowledgments}

\appendix 

\begin{widetext}

\section{Derivation of analytic Hubbard-type interference formulas for two particles}
\label{a1}

Here we illustrate in detail the derivation of the analytic interference formulas for $N=2$ particles, 
allowing a rather immediate generalization to more complex cases with $N > 2$ particles. For 
this analytic modeling, we assume that the spatial part of the orbital of the $j$th particle is 
approximated by a displaced Gaussian function (localized at a position $d_j$),
\begin{equation}
\psi_{j}(x)=\frac{1}{(2\pi)^{1/4}\sqrt{s}}\exp \left(-\frac{(x-d_j)^2}{4s^2}\right),
\label{psix}
\end{equation}
where $s$ denotes the width of the Gaussian functions. The single-particle orbital $\psi_j (k)$ in the 
momentum Hilbert space is given by the Fourier transform of $\psi_j (x)$, namely 
$\psi_j (k) = (1/\sqrt{2\pi})\int_{-\infty}^{\infty}\psi_j (x)\exp (ikx) dx$. Performing this Fourier 
transform, one finds
\begin{equation}
\psi_{j}(k)=\frac{2^{1/4}\sqrt{s}}{\pi^{1/4}}\exp (-k^2s^2)\exp (id_j k),
\label{psik}
\end{equation}

In our previous paper \cite{yann17}, we focused on well localized particles within each well (neglecting
the possibility of double occupancy in each well), a condition 
that is satisfied for strong repulsion. Here we are interested in an 
analytical model for all interaction strengths, allowing for double occupancy. We therefore consider the
more general case of the two-site Hubbard model instead of the Heisenberg model (as was done in Ref.\
\cite{yann17}). The two particles are localized at two different wells, at positions $d_1<0$ and $d_2>0$, 
which together with the spin yields four possible spin eigenfunctions 
$\ket{\circ,\uparrow\downarrow},\ket{\downarrow,\uparrow},\ket{\uparrow,\downarrow}$, and 
$\ket{\uparrow\downarrow,\circ}$. These spin eigenfunctions form a complete many-body base for the 
diagonalization of the fermionic Hubbard Hamiltonian,
\begin{align}
H=-\sum_{\sigma}\left(\hat{c}_{1,\sigma}^{\dagger}\hat{c}_{2,\sigma}+
\hat{c}_{2,\sigma}^{\dagger}\hat{c}_{1,\sigma}\right)+
{\cal U} \sum_{i=1}^2\hat{n}_{i\uparrow}\hat{n}_{i\downarrow},
\label{hubh}
\end{align}
where $\sigma$ sums over the up ($\uparrow$) and down ($\downarrow$) spins. The ratio $\mathcal{U}=U/t$,
where $U$ and $t$ are the one-site repulsion and the nearest-neighbor hopping parameters. The energies are 
expressed in units of $t$. 

There are many equivalent ways of writing the Hubbard model basis in the second-quantization
formalism, and throughout this paper we use the following convention
\begin{align}
b_1&=\hat{c}_{2\uparrow}^\dagger \hat{c}_{2\downarrow}^\dagger\ket{}=\ket{RR}=\ket{\circ,\uparrow\downarrow},
\label{eq:basisfun1}\\
b_2&=\hat{c}_{1\downarrow}^\dagger\hat{c}_{2\uparrow}^\dagger\ket{}=\ket{RL}=\ket{\downarrow,\uparrow},
\label{eq:basisfun2}\\
b_3&=\hat{c}_{1\uparrow}^\dagger\hat{c}_{2\downarrow}^\dagger\ket{}=\ket{LR}=\ket{\uparrow,\downarrow},
\label{eq:basisfun3}\\
b_4&=\hat{c}_{1\uparrow}^\dagger\hat{c}_{1\downarrow}^\dagger\ket{}=\ket{LL}=\ket{\uparrow\downarrow,\circ}.
\label{eq:basisfun4}
\end{align}

In the third column above, the spin of the particle is not denoted explicitly. In this case the
following mnemonic rule is helpful: the spin-up particle is always written first inside the ket
$\ket{\cdots}$

For a small number of particles the Hubbard Hamiltonian can be exactly diagonalized (for instance using 
SNEG \cite{sneg}). For $S_z=0$, a general solution of the two-site Hubbard Hamiltonian using the
aforementioned second-quantization basis is of the form
\begin{align}
\Phi=a(\mathcal{U})\ket{\circ,\uparrow\downarrow}+b(\mathcal{U})\ket{\downarrow,\uparrow}+
c(\mathcal{U})\ket{\uparrow,\downarrow}+d(\mathcal{U})\ket{\uparrow\downarrow,\circ}.
\end{align}

The coefficients $a({\cal U}),\ldots,d({\cal U})$ of course satisfy the constraint that $\Phi$ is 
normalized. Naturally, such a Hubbard-model solution yields the wave function in second quantization form. 
Our modification aims at including the spatial component of the wave-function, by associating 
each basis ket $b_i$, $i=1,\ldots,4$, with a determinant of spin orbitals 
$\psi_{j,\sigma}(x)=\psi_j(x)\sigma$, where $\sigma$ here represents the spin. When the spin orbitals
are localized on the left or right well, they will also be denoted as $|L\rangle$ or $|R\rangle$,
respectively. The corresponding determinants ${\cal D}$ to each basis ket are (the tilde indicates the 
incorporation of the space orbitals) 
\begin{align}
\ket{\widetilde{RR}}&=\mathcal{D}_{\ket{\circ,\uparrow\downarrow}}(x_1,x_2)\nonumber\\&=\frac{1}{\sqrt{2!}}(\psi_{2\uparrow}(x_1) 
\psi_{2\downarrow}(x_2)-\psi_{2\uparrow}(x_2)\psi_{2\downarrow}(x_1))
\end{align}
\begin{align}
\ket{\widetilde{RL}}&=\mathcal{D}_{\ket{\downarrow,\uparrow}}(x_1,x_2)\nonumber\\&=
\frac{1}{\sqrt{2!}}(\psi_{1\downarrow}(x_1) \psi_{2\uparrow}(x_2)-
\psi_{1\downarrow}(x_2)\psi_{2\uparrow}(x_1))
\end{align}
\begin{align}
\ket{\widetilde{LR}}&=\mathcal{D}_{\ket{\uparrow,\downarrow}}(x_1,x_2)\nonumber\\&=
\frac{1}{\sqrt{2!}}(\psi_{1\uparrow}(x_1) \psi_{2\downarrow}(x_2)-
\psi_{1\uparrow}(x_2)\psi_{2\downarrow}(x_1))
\end{align}
\begin{align}
\ket{\widetilde{LL}}&=\mathcal{D}_{\ket{\uparrow\downarrow,\circ}}(x_1,x_2)\nonumber\\&=\frac{1}{\sqrt{2!}}(\psi_{1\uparrow}(x_1) 
\psi_{1\downarrow}(x_2)-\psi_{1\uparrow}(x_2)\psi_{1\downarrow}(x_1))
\end{align}

We can therefore write the full wave function, including the space and spin parts, as 
\begin{align}
\begin{split}
\Phi(x_1,x_2)=&a(\mathcal{U})\mathcal{D}_{\ket{\circ,\uparrow\downarrow}}(x_1,x_2)+
b(\mathcal{U})\mathcal{D}_{\ket{\downarrow,\uparrow}}(x_1,x_2)+\\
&c(\mathcal{U})\mathcal{D}_{\ket{\uparrow,\downarrow}}(x_1,x_2)+
d(\mathcal{U})\mathcal{D}_{\ket{\uparrow\downarrow,\circ}}(x_1,x_2),
\end{split}
\end{align}
where the coefficients are in general dependent on the interwell distance $d=d_1-d_2$ and the width $s$. 

We can now use the wave function $\Phi(x_1,x_2)$, together with the formulas described in the main paper
[see Eqs.\ (3) and (4) therein], to obtain the two-particle correlation expressions in real and 
momentum space [see Eqs.\ (5)-(8) in the main paper]. The integrations associated with the Fourier 
transforms can be carried out with the help of the MATHEMATICA algebraic computer language \cite{math}.\\
~~~\\

\begin{figure*}[t]
\centering\includegraphics[width=10.5cm]{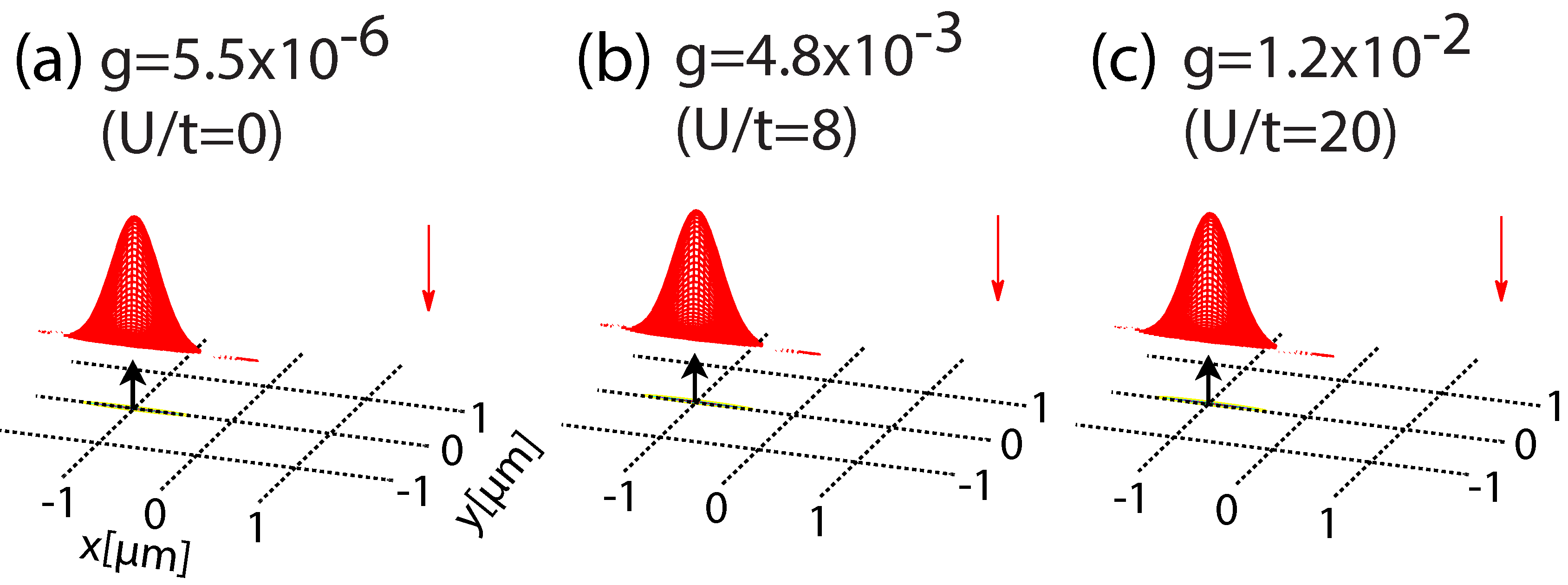}
\caption{This figure shows spin-resolved conditional probability densities (SR-CPDs) for two particles in 
the second excited state in a double well. We plot the SR-CPD for three different interaction strengths 
from $g=5.5\; 10^{-6} \hbar \omega l_0^2$ (corresponding to $U/t=0$) to $g=1.2\; 10^{-2} 
\hbar \omega l_0^2$ (corresponding to $U/t=20$). The black up-arrow represents the fixed position of the
spin-up particle in the plane of the grid. The red down-arrow indicates that we are calculating the 
resulting occupation probability (density) for a spin down particle; see the plotted red-color surface. 
As is apparent from the figure, the red-cplored probability surface is directly situated on top of the 
black (spin-up) fixed point for all interaction strengths. This indicates strong double occupancy. The 
second well of the double well at $d_2=d/2>0$ is practically unoccupied. This double occupancy is what 
allows us to extract the Hubbard on-site interaction parameter $U$ from the energy level of the second 
excited CI state. The parameters for the double wells are: $\hbar\omega=\hbar\omega_x=15$ kHz, 
$\hbar\omega_y=1500$ kHz, $\epsilon_b$=0.5 ($V_b=33.5$ kHz). The interwell distance is $d=2$ $\mu$m.}
\label{figs1}
\end{figure*}

\section{Extraction of Hubbard-model parameters from the CI calculation}
\label{a2}

In order to compare our analytical model with the CI results it is important to relate the interparticle
interaction strength $g$ [see Eq.\ (2) in the main paper] with the Hubbard parameter $U$, and to extract the
value of the hopping parameter $t$ from the single-particle energy spectrum associated with the external 
confining potential. Given the single-particle spectrum, the value of $t$ can be extracted as 
$t=(e_2-e_1)/2$ where $e_1$ and $e_2$ are the ground and first-excited single-particle energies,
respectively. This can be directly inferred from the tight-binding limit (setting $U=0$). 

In order to determine $U$ from the CI, we first take a close look at the Hubbard-model energy levels and 
their properties. An exact diagonalization of the Hubbard Hamiltonian shows that the second excited state 
energy $E_3(U)$ is directly proportional to $U$ with $E_{3}(U)=U+2t+E_1(0)$, where $E_1(0)$ is the 
non-interacting ground state energy. For non-interacting ($U=0$) particles, the energy of the second 
excited state is therefore simply given as $E_3(0)=2t+E_1(0)$. Consequently one can extract the parameter 
$U$ directly from the difference between the non-interacting and interacting second excited-state energy 
$U=E_3(U)-E_3(0)$. This is a trivial result within the Hubbard model, but it also applies for our CI 
calculations. 

In order to verify that $U$ can be determined by using the corresponding energy difference from our CI 
spectrum, i.e., $U=E_3^{\rm CI}(g)-E_3^{\rm CI}(0)$, we look at the properties of the second-excited CI 
state. In the Hubbard model the second excited state is given as $(\ket{LL}-\ket{RR})/\sqrt{2}$, containing
only doubly occupied sites (as we would expect since $U$ represents the on-site interaction energy). It is 
easily verified via conditional probability distributions (CPDs) \cite{yann07,li09,yann15,yann16} that, in 
analogy with the Hubbard-model case, the second excited CI state consists solely of doubly occupied wells; 
see Fig.\ \ref{figs1}. We therefore proceed to determine 
$U$ using $U=E_3^{\rm CI}(g)-E_3^{\rm CI}(0)$. Afterwards we compare the CI and Hubbard energy levels using
values for $U$ obtained from the CI in this way and find very good agreement between the CI spectrum and 
the Hubbard model spectrum [see Figs.\ 2(e) and 3(a) in main paper], validating our approach for
extracting $U$ from the CI calculation.\\
~~~~\\

\begin{figure*}[t]
\centering\includegraphics[width=10.5cm]{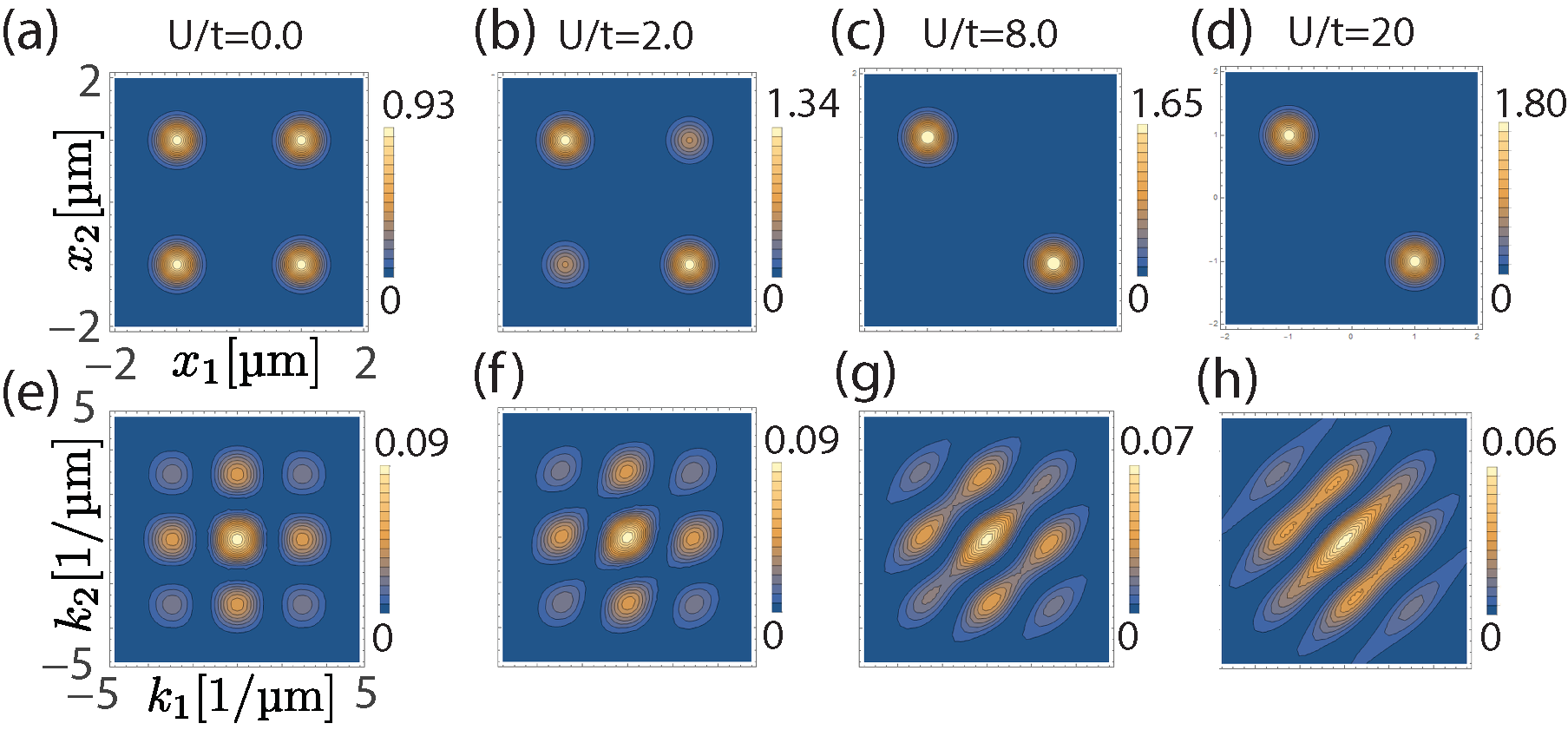}
\caption{The analytic Hubbard-model ground-state (singlet) space (a-d) and momentum [(e-h), see Eq.\ (5) 
in main paper] two-particle correlation maps for two ultracold fermions in a double well, as a function of 
the Hubbard interaction strength $U/t$. The interwell distance is $d=2$ $\mu$m and the width of the 
displaced Gaussian functions is $s=0.2$ $\mu$m.}
\label{figs2}
\end{figure*}

\begin{figure}[t]
\centering\includegraphics[width=8.cm]{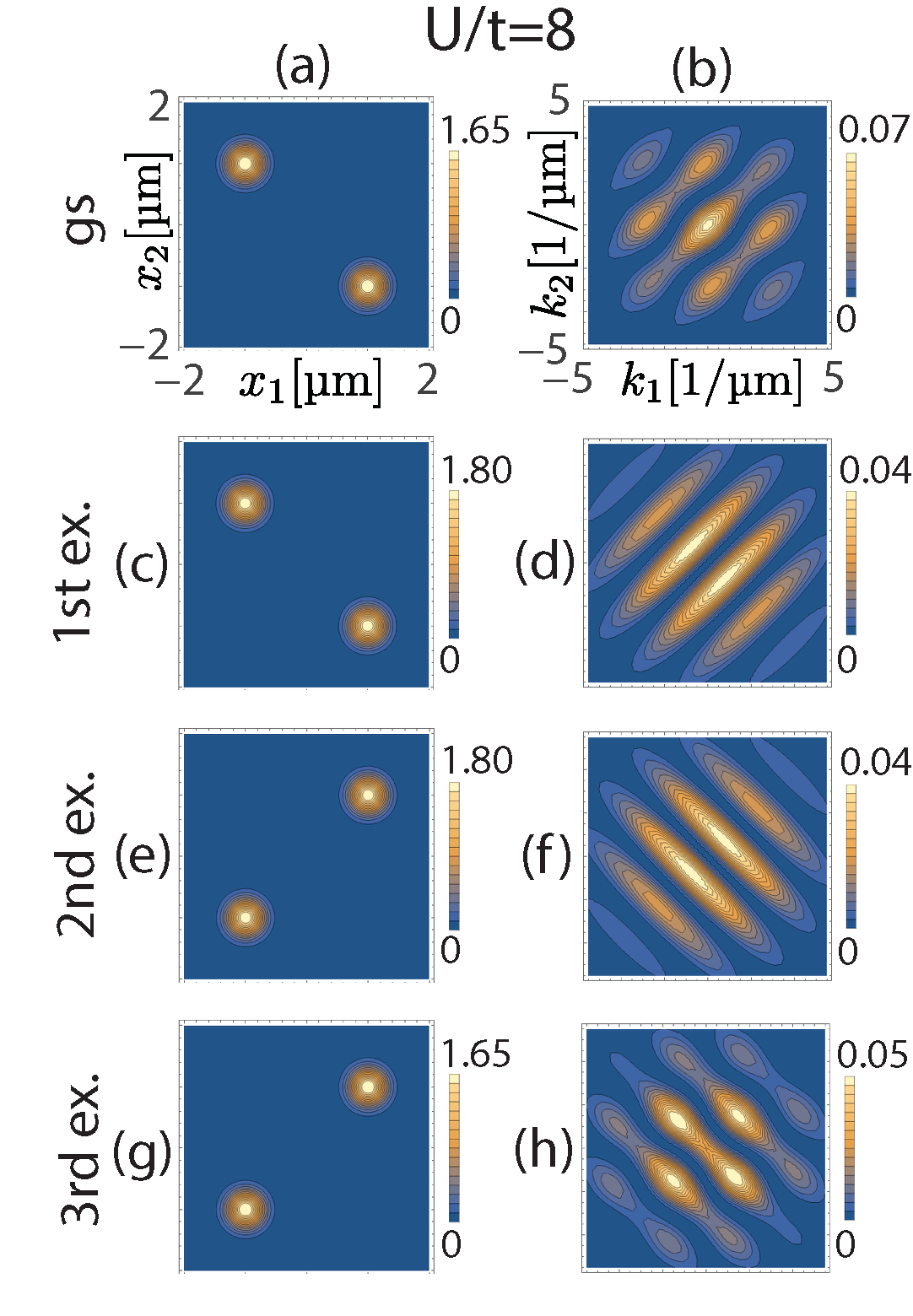}
\caption{Hubbard-model analytic two-particle correlation maps for the ground state and the three lowest 
excited states (as marked in the figure) of two ultracold fermions in a double well, calculated for an 
intermediate {\it positive} value of the Hubbard interaction strength $U/t = 8$. (a,c,e,g) Two-particle 
spatial correlations maps. (b,d,f,h) Two-particle momentum correlation maps according to Eqs.\ (5)$-$(8) 
in the main paper. The interwell distance is $d=2$ $\mu$m and the width of the displaced Gaussian functions
is $s=0.2$ $\mu$m.}
\label{figs3}
\end{figure}

\begin{figure}[t]
\centering\includegraphics[width=8.cm]{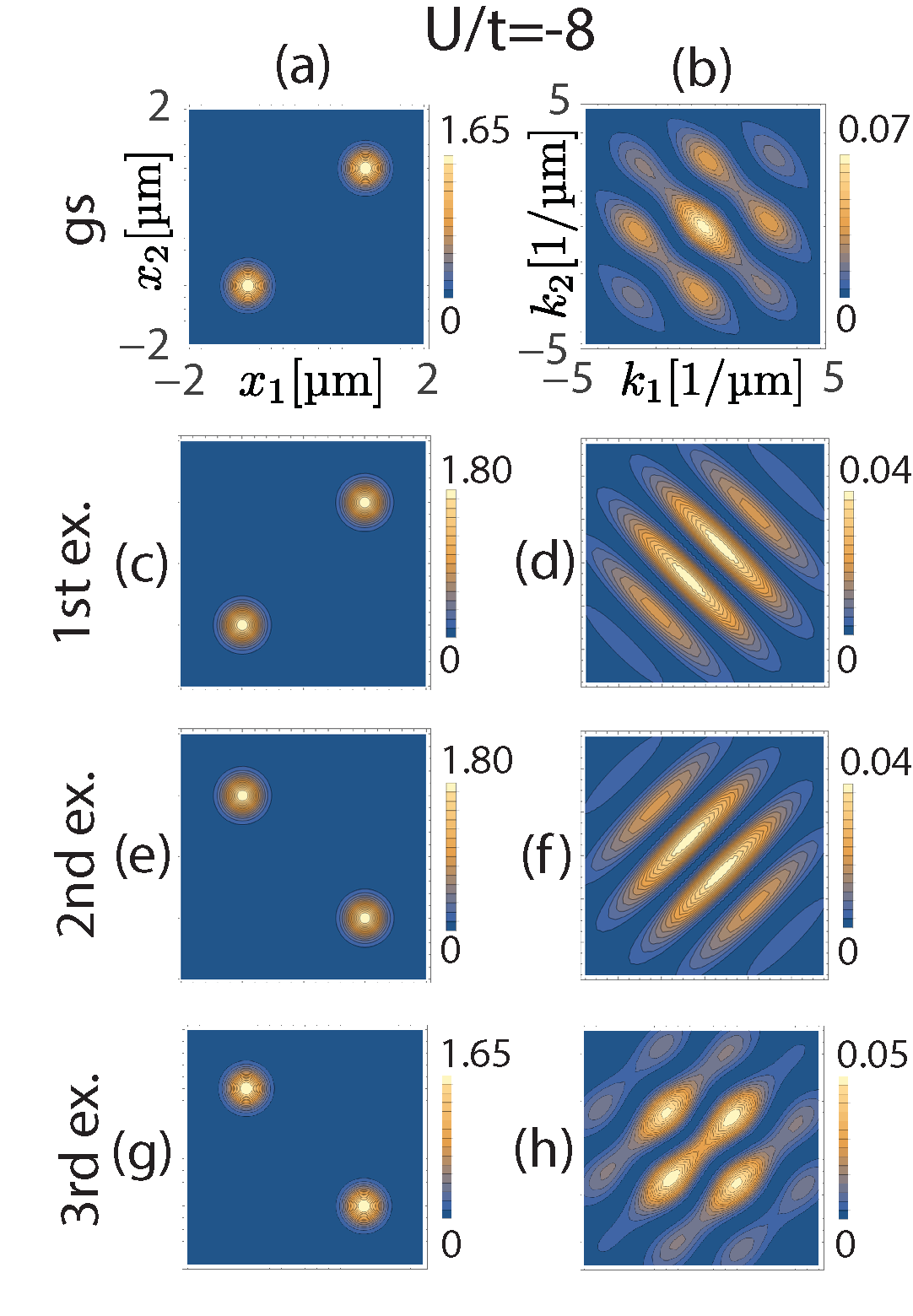}
\caption{Hubbard-model analytic two-particle correlation maps for the ground state and the three lowest 
excited states (as marked in the figure) of two ultracold fermions in a double well, calculated for an 
intermediate {\it negative} value of the Hubbard interaction strength $U/t = -8$. (a,c,e,g) Two-particle 
spatial correlations maps. (b,d,f,h) Two-particle momentum correlation maps according to Eqs.\ (5)$-$(8) in
the main paper. The interwell distance is $d=2$ $\mu$m and the width of the displaced Gaussian functions is 
$s=0.2$ $\mu$m.}
\label{figs4}
\end{figure}

\section{Additional figures portraying Hubbard-model two-particle momentum correlation maps}
\label{a3}

In order to further highlight the extent to which our modified-solutions Hubbard model reproduces the
microscopic CI two-particle space and momentum correlations, we display here three additional Figures 
\ref{figs2}, \ref{figs3}, and \ref{figs4}. Both space and momentum correlation maps in Fig. \ref{figs2} 
should be compared with the corresponding CI ones in the bottom row (steeper confinement with 
$\hbar \omega=15$ kHz) of Fig.\ 1 in the main paper. The momentum correlation maps in Fig.\ \ref{figs3} 
for the repulsive case with $U/t = 8$ should be compared with those CI ones in Figs.\ 3(b,c,e) in the main 
paper (due to the contrast, the outer fringes in Figs.\ 3(b,c) are better seen after one enlarges these 
figure panels). The corresponding results for the attractive case with $U/t = -8$ are shown for 
completeness in Fig.\ \ref{figs4}.\\
~~~~~\\

\section{The 2nd-order (two-body) density matrices derived in the Hilbert space of the modified-solutions
Hubbard model and their relation to the correlation maps}
\label{a4}

\subsection{Solution of the two-site two-particle Hubbard model}
\label{a41}

Here we outline the solution of the two-site Hubbard model with two spin $1/2$ fermions. The Hubbard Hamiltonian in second quantization is given in Eq.\ (\ref{hubh}).
We remind that $U$ in Eq.\ (\ref{hubh}) is the on-site interaction, $t$ is the tunneling parameter and 
$\hat{n}_{i\sigma}$ is the number operator at site $i$ for spin $\sigma$. For convenience we repeat here 
our definition of the Hubbard model basis functions:
\begin{align}
b_1&=\hat{c}_{2\uparrow}^\dagger \hat{c}_{2\downarrow}^\dagger\ket{}=\ket{RR}=\ket{\circ,\uparrow\downarrow},
\label{eq:basisfun1v}\\
b_2&=\hat{c}_{1\downarrow}^\dagger\hat{c}_{2\uparrow}^\dagger\ket{}=\ket{RL}=\ket{\downarrow,\uparrow},
\label{eq:basisfun2v}\\
b_3&=\hat{c}_{1\uparrow}^\dagger\hat{c}_{2\downarrow}^\dagger\ket{}=\ket{LR}=\ket{\uparrow,\downarrow},
\label{eq:basisfun3v}\\
b_4&=\hat{c}_{1\uparrow}^\dagger\hat{c}_{1\downarrow}^\dagger\ket{}=\ket{LL}=\ket{\uparrow\downarrow,\circ},
\label{eq:basisfun4v}
\end{align}
where $L$ and $R$ represent site $1$ and $2$ respectively. There are many equivalent notations for these basis functions in the literature and we have listed three of them in Eqs.\ 
(\ref{eq:basisfun1})-(\ref{eq:basisfun4}). In the following we will use the $L,R$ 
notation. The basis set in Eqs.\ (\ref{eq:basisfun1})-(\ref{eq:basisfun4})
spans the Hilbert space of the 2-site 2-particle Hubbard model and the resulting Hubbard matrix is:
\begin{align}
\mathscr{H}=
  \begin{bmatrix}
    U&t&-t&0\\
    t&0&0&t\\
    -t&0&0&-t\\
    0&t&-t&U
  \end{bmatrix}.
\end{align}
\noindent Diagonalization of this Hamiltonian yields the eigenenergies:
\begin{align}
E_1&=\frac{1}{2} \left(U-\sqrt{16 t^2+U^2}\right),\\
E_2&=0,\\
E_3&=U,\\
E_4&=\frac{1}{2} \left(U+\sqrt{16 t^2+U^2}\right).
\end{align}

  
\noindent The eigenfunctions corresponding to these eigenvalues are:

\begin{align}
\Phi_{1}&=A(U,t)\ket{RR}+B(U,t)\ket{RL}-B(U,t)\ket{LR}+A(U,t)\ket{LL} \nonumber\\
&=A(U,t)(\ket{RR}+\ket{LL})+B(U,t)(\ket{RL}-\ket{LR}) \nonumber \\
&=A(U,t)\sqrt{2}\ket{B_4}-B(U,t)\sqrt{2}\ket{B_1},
\label{eq:hubbardgsvariant2}\\
\Phi_{2}&=\frac{1}{\sqrt{2}}\left(\ket{RL}+\ket{LR}\right)=\ket{B_2},
\label{eq:hubbard1stvariant2}\\
\Phi_{3}&=\frac{1}{\sqrt{2}}\left(\ket{LL}-\ket{RR}\right)=\ket{B_3},
\label{eq:hubbard2ndvariant2}\\
\Phi_{4}&=C(U,t)\ket{RR}+D(U,t)\ket{RL}-D(U,t)\ket{LR}+C(U,t)\ket{LL}\nonumber \\
&=C(U,t)(\ket{RR}+\ket{LL})+D(U,t)(\ket{RL}-\ket{LR}) \nonumber \\
&=C(U,t)\sqrt{2}\ket{B_4}-D(U,t)\sqrt{2}\ket{B_1},\label{eq:hubbard3rdvariant2}
\end{align}

\noindent where 

\begin{align}
A(U,t)&=\left. 1\middle/\sqrt{\frac{\left(\sqrt{16 t^2+U^2}+U\right)^2}{8 t^2}+2} \right. ,\\
B(U,t)&=\left. -\frac{\sqrt{16 t^2+U^2}+U}{4 t}\middle/
\sqrt{\frac{\left(\sqrt{16 t^2+U^2}+U\right)^2}{8 t^2}+2}\right.,\\
C(U,t)&=\left. 1\middle/ \sqrt{\frac{\left(U-\sqrt{16 t^2+U^2}\right)^2}{8 t^2}+2}\right.,\\
D(U,t)&=\left. \frac{-U+\sqrt{16 t^2+U^2}}{4 t}\middle/ 
\sqrt{\frac{\left(U-\sqrt{16 t^2+U^2}\right)^2}{8 t^2}+2}\right.,
\end{align}

\noindent and $\ket{B_1},\ket{B_2},\ket{B_3},\ket{B_4}$ are the four Bell states:

\begin{align}
\ket{B_1}&=\frac{1}{\sqrt{2}}(\ket{LR}-\ket{RL})\\
\ket{B_2}&=\frac{1}{\sqrt{2}}(\ket{LR}+\ket{RL})\\
\ket{B_3}&=\frac{1}{\sqrt{2}}(\ket{LL}-\ket{RR})\\
\ket{B_4}&=\frac{1}{\sqrt{2}}(\ket{LL}+\ket{RR}).
\end{align}  
  
\noindent Writing the Hubbard model solutions in this form has the advantage that it reveals the simple structure of the Hubbard Hamiltonian in the four Bell-states basis, i.e.,

\begin{align}
\mathscr{H}=
  \mathscr{H}=
  \begin{bmatrix}
    0&0&0&-2t\\
    0&0&0&0\\
    0&0&U&0\\
    -2t&0&0&U
  \end{bmatrix}.
\end{align}\\
~~~~\\ 
We note that for two fermions, one can form two additional Bell states by adding and subtracting the
$S=1,\;S_z=1$ ($\ket{\uparrow,\uparrow}$) and $S=1,\;S_z=-1$ ($\ket{\downarrow,\downarrow}$) 
triplet states \cite{schl01,schl02}. These two Bell states, however, do not conserve the total spin, and 
thus they are not CI eigenstates.   
  
\subsection{Calculation of the second-order two-body density matrix}
\label{a42}

For each state of the Hubbard model ($\Phi_{1},\Phi_{2},\Phi_{3},\Phi_{4}$, denoted in general 
as $\Phi$) one can obtain the second-order density matrix as 
$\rho_{\rm Hub}^{S,S_z}=\ket{\Phi}\bra{\Phi}$, which can be written in the $L$, $R$ basis as

\begin{align}
\rho_{\rm Hub}^{S,S_z}=\sum_{i,j,k,l=L,R}\rho_{ijkl}^{\text{Hub},S,S_z}\ket{i\;j}\bra{k\;l}.
\end{align}

\noindent In order to obtain the spatial second-order density matrix (and subsequently the second-order spatial correlation function) from the Hubbard model density matrix we define an operator that associates single particle spatial wavefunctions [$\psi$, see, e.g., Eq.\ (\ref{psix})] with the $L,R$ 
basis, as 

\begin{align}
\mathcal{O}_s=\sum_{i,j,k,l=L,R}\psi_{i\uparrow}(x_1)\psi_{j\downarrow}(x_2)\psi_{k\uparrow}^\dagger(x_1')\psi_{l\downarrow}^\dagger(x_2')\hat{c}_{1\uparrow}\hat{c}_{2\downarrow}\hat{c}^\dagger_{1\uparrow}\hat{c}_{2\downarrow}^\dagger.
\end{align} 

\noindent The spatial second-order density matrix can then be obtained as the expectation value of this 
operator

\begin{align}
\mathcal{G}_{\rm Hub}^{S,S_z}(x_1,x_2,x_1',x_2')=\langle \Phi | \mathcal{O}_s |\Phi \rangle=
\Tr[\rho_{\rm Hub}^{S,S_z}\;\mathcal{O}_s],
\end{align}

\noindent which yields

\begin{align}
\mathcal{G}_{\rm Hub}^{S,S_z}(x_1,x_2,x_1',x_2')=\sum_{i,j,k,l=L,R}\rho_{ijkl}^{\text{Hub},S,S_z}\psi_{i\uparrow}(x_1)\psi_{j\downarrow}(x_2)\psi_{k\uparrow}^\dagger(x_1')\psi_{l\downarrow}^\dagger(x_2').
\end{align}

\noindent The second-order momentum density matrix is obtained through Fourier transform

\begin{align}
\begin{split}
\mathcal{G}_{\rm Hub}^{S,S_z}(k_1,k_2,k_1',k_2')&=\frac{1}{4\pi^2}\int_{-\infty}^\infty e^{-k_1 x_1}dx_1 \int_{-\infty}^\infty e^{-k_2 x_2}dx_2 \int_{-\infty}^\infty e^{k_1' x_1'}dx_1' \\&\;\;\;\;\;\;\;\;\; \int_{-\infty}^\infty e^{k_2' x_2'}dx_2'                         \sum_{i,j,k,l=L,R}\rho_{ijkl}^{\text{Hub},S,S_z}\psi_i(x_1)\psi_j(x_2)\psi_k^\dagger(x_1')\psi_l^\dagger(x_2'),\\&=\sum_{i,j,k,l=L,R}\rho_{ijkl}^{\text{Hub},S,S_z}\psi_{i\uparrow}(k_1)\psi_{j\downarrow}(k_2)\psi_{k\uparrow}^\dagger(k_1')\psi_{l\downarrow}^\dagger(k_2').
\end{split}
\end{align}

\noindent To proceed we use single particle Gaussian wavefunctions for the left and right wells, where $L,R$ indicate that the real-space Gaussian wavefunction ($\psi$) is localized in the left ($\psi_L$) and right ($\psi_R$) well respectively. The real-space displaced Gaussian function was given in Eq.\ (\ref{psix});
($d_j<0$ corresponds to $L$, $d_j>0$ corresponds to $R$) and $s$ is the Gaussian width. Its fourier transform was
given in Eq.\ (\ref{psik}).

\noindent Using these Gaussian single-particle wavefunctions, the second-order momentum density-matrix elements can 
be calculated explicitly,

\begin{align}
\eta_{ijkl}^{\text{Hub},S,S_z}(k_1,k_2,k_1',k_2')=\rho_{ijkl}^{\text{Hub},S,S_z}\psi_{i\uparrow}(k_1)\psi_{j\downarrow}(k_2)\psi_{k\uparrow}^\dagger(k_1')\psi_{l\downarrow}^\dagger(k_2').
\end{align}

\noindent This allows us to write the second-order momentum density matrix as

\begin{align}
\mathcal{G}_{\rm Hub}^{S,S_z}(k_1,k_2,k_1',k_2')=\sum_{i,j,k,l=L,R}\eta_{ijkl}^{\text{Hub},S,S_z}(k_1,k_2,k_1',k_2').
\end{align} 

\noindent For a physical interpretation and for the creation of the second-order momentum correlation maps 
we are interested only in the diagonal elements of the second-order momentum density matrix, 
which are given as 

\begin{align}
\mathcal{G}_{\rm Hub}^{S,S_z}(k_1,k_2)&=\sum_{i,j,k,l=L,R}\eta_{ijkl}^{\text{Hub},S,S_z}(k_1,k_2),
\label{eq:2ndordercorrelationfunsum}
\end{align}
\noindent with
\begin{align}
\mathcal{G}_{\rm Hub}^{S,S_z}(k_1,k_2)&\equiv\mathcal{G}_{\rm Hub}^{S,S_z}(k_1,k_2,k_1,k_2),\\
\eta_{ijkl}^{\text{Hub},S,S_z}(k_1,k_2)&\equiv\eta_{ijkl}^{\text{Hub},S,S_z}(k_1,k_2,k_1,k_2).
\end{align}

\noindent When evaluating this expression one needs to account for the orthogonality of the spins. The function $\mathcal{G}_{\rm Hub}^{S,S_z}(k_1,k_2)$ is termed second-order (two-body) momentum correlation function. One can obtain the spin-resolved version by only selecting terms with a certain spin configuration. Alternatively, the spin-unresolved version can be obtained by taking all the spin terms into account. In the special case of a two-particle second-order correlation function, both the spin resolved and the spin unresolved versions are identical (for a given spin-projection) apart from an overall factor. Expressing $\mathcal{G}_{\rm Hub}^{S,S_z}(k_1,k_2)$ using the $\eta_{ijkl}^{\text{Hub},S,S_z}(k_1,k_2)$ elements has the advantage that the $\eta_{ijkl}^{\text{Hub},S,S_z}(k_1,k_2)$ clearly show the interference terms that correspond to the individual entries in the Hubbard model density matrix $\rho_{\rm Hub}^{S,S_z}$.  These elements can be readoff directly from the matrices given in Appendices \ref{a43}, \ref{a44}, \ref{a45}, and \ref{a46}.\\ 
~~~~\\
\newpage 

\subsection{Ground state}
\label{a43}

Using $\mathcal{U}=U/t$ and  $\mathcal{Q}(\mathcal{U})=\sqrt{16+\mathcal{U}^2}+\mathcal{U}$ the Hubbard model two-body density matrix is given by

\begin{align}
\arraycolsep=1.4pt\def\arraystretch{2.2}
\rho_{\rm Hub}^{S=0,S_z=0}&=\frac{1}{\mathcal{Q}(\mathcal{U}) \mathcal{U}+16} \;\;\;
\begin{blockarray}{ccccc}
 LL & LR & RL & RR & \\ 
\begin{block}{(cccc)c}
 4 & ~~\mathcal{Q}(\mathcal{U})~~ & \mathcal{Q}(\mathcal{U}) & 4 & ~~LL \\
 & ~~\frac{\mathcal{Q}(\mathcal{U})\mathcal{U}}{2}+4~~ 
& ~~\frac{\mathcal{Q}(\mathcal{U})\mathcal{U}}{2}+4~~ 
& \mathcal{Q}(\mathcal{U}) & ~~LR \\
 \BAmulticolumn{2}{c}{\multirow{2}{*}{{\LARGE h.c.}}} & 
~~\frac{\mathcal{Q}(\mathcal{U})\mathcal{U}}{2}+4~~ & \mathcal{Q}(\mathcal{U}) & ~~RL \\
 & & & 4 & ~~RR \\
\end{block}
\end{blockarray},\\
&=\;\;\;\begin{blockarray}{ccccc}
 LL & LR & RL & RR & \\ 
\begin{block}{(cccc)c}
 A(\mathcal{U})^2 & -A(\mathcal{U})B(\mathcal{U}) & -A(\mathcal{U})B(\mathcal{U}) & A(\mathcal{U})^2 & ~~LL \\
 & B(\mathcal{U})^2 & B(\mathcal{U})^2 &  -A(\mathcal{U})B(\mathcal{U}) & ~~LR \\
 \BAmulticolumn{2}{c}{\multirow{2}{*}{{\LARGE h.c.}}} & B(\mathcal{U})^2 &  -A(\mathcal{U})B(\mathcal{U}) & ~~RL \\
 & & & A(\mathcal{U})^2 & ~~RR \\
\end{block}
\end{blockarray}.
\end{align}
 
\noindent Note that $\rho_{\rm Hub}^{S=0,S_z=0}$ for the ground state as well as for the excited states 
(see Appendices \ref{a44}, \ref{a45}, and \ref{a46} below) are idempotent. 
Including the Fourier transformed wave functions we obtain

{\footnotesize
\begin{align}
& \eta_{\rm Hub}^{S=0,S_z=0}(k_1,k_2)= \nonumber\\
\arraycolsep=1.4pt\def\arraystretch{2.2}
&\frac{2 s ^2 e^{-2 s ^2 \left(k_1^2+k_2^2\right)}}{\pi  (\mathcal{Q}(\mathcal{U}) \mathcal{U}+16)} \;\;\;
\begin{blockarray}{ccccc}
 LL & LR & RL & RR & \\ 
\begin{block}{(cccc)c}
 4 & e^{-2 i d k_2} \mathcal{Q}(\mathcal{U}) & e^{-2 i d k_1} \mathcal{Q}(\mathcal{U}) & 4 e^{-2 i d (k_1+k_2)} & ~~LL \\
 & \frac{\mathcal{Q}(\mathcal{U})\mathcal{U}}{2}+4 & ~~\frac{1}{2} e^{-2 i d (k_1-k_2)} (\mathcal{Q}(\mathcal{U}) \mathcal{U}+8)~~ & e^{-2 i d k_1} \mathcal{Q}(\mathcal{U}) & ~~LR \\
  \BAmulticolumn{2}{c}{\multirow{2}{*}{{\LARGE h.c.}}} & \frac{\mathcal{Q}(\mathcal{U})\mathcal{U}}{2}+4 & e^{-2 i d k_2} \mathcal{Q}(\mathcal{U}) & ~~RL \\
 & & & 4 & ~~RR \\
\end{block}
\end{blockarray}.
\label{eq:gsmatrixwithmomentum}
\end{align}
}

\noindent Using Eq.\ (\ref{eq:2ndordercorrelationfunsum}) and the second-order momentum matrix in 
Eq.\ (\ref{eq:gsmatrixwithmomentum}), one can obtain the two-body ground state momentum correlation 
function [see Eq.\ (5) in the main paper]. Similarly the two-body momentum correlation functions for the 
excited states [see Eqs.\ (6)-(8) in the main paper] can be obtained through the use of the matrices 
given in Appendices \ref{a44}, \ref{a45}, and \ref{a46} below.

\newpage

\subsection{1st excited state}
\label{a44}

$\rho_{\rm Hub}^{S=1,S_z=0}$ and $\eta_{\rm Hub}^{S=1,S_z=0}(k_1,k_2)$ for the first excited state of the 
Hubbard Hamiltonian [see Eq.\ (\ref{eq:hubbard1stvariant2})] are given by:
\begin{align}
\arraycolsep=1.4pt\def\arraystretch{2.2}
\rho_{\rm Hub}^{S=1,S_z=0}=\frac{1}{2} \;\;\;
\begin{blockarray}{ccccc}
 LL & LR & RL & RR & \\ 
\begin{block}{(cccc)c}
 0 & 0 & 0 & 0 & ~~LL \\
 & 1 & -1 & 0 & ~~LR \\
  \BAmulticolumn{2}{c}{\multirow{2}{*}{{\LARGE h.c.}}} & 1 & 0 & ~~RL \\
 & & & 0 & ~~RR\\
\end{block}
\end{blockarray},
\end{align}
 
\begin{align}
\arraycolsep=1.4pt\def\arraystretch{2.2}
\eta_{\rm Hub}^{S=1,S_z=0}(k_1,k_2)=\frac{4 s^2 e^{-2 s^2 \left(k_1^2+k_2^2\right)}}{\pi } \;\;\;
\begin{blockarray}{ccccc}
 LL & LR & RL & RR & \\ 
\begin{block}{(cccc)c}
 0 & 0 & 0 & 0 & ~~LL \\
 & \frac{1}{4} & -\frac{1}{4} e^{-2 i d (k_1-k_2)} & 0 & ~~LR \\
  \BAmulticolumn{2}{c}{\multirow{2}{*}{{\LARGE h.c.}}} & \frac{1}{4} & 0 & ~~RL \\
 & & & 0 & ~~RR \\
\end{block}
\end{blockarray}.
\end{align}

\newpage  

\subsection{2nd excited state}
\label{a45}

$\rho_{\rm Hub}^{S=0,S_z=0}$ and $\eta_{\rm Hub}^{S=0,S_z=0}(k_1,k_2)$ for the second excited state of the 
Hubbard Hamiltonian [see Eq.\ (\ref{eq:hubbard2ndvariant2})] are given by:

\begin{align}
\arraycolsep=1.4pt\def\arraystretch{2.2}
\rho_{\rm Hub}^{S=0,S_z=0}=\frac{1}{2} \;\;\;
\begin{blockarray}{ccccc}
 LL & LR & RL & RR & \\ 
\begin{block}{(cccc)c}
 1 & 0 & 0 & -1 & ~~LL \\
 & 0 & 0 & 0 & ~~LR \\
  \BAmulticolumn{2}{c}{\multirow{2}{*}{{\LARGE h.c.}}} & 0 & 0 & ~~RL \\
 & & & 1 & ~~RR\\
\end{block}
\end{blockarray},
\end{align}
 
\begin{align}
\arraycolsep=1.4pt\def\arraystretch{2.2}
\eta_{\rm Hub}^{S=0,S_z=0}(k_1,k_2)=\frac{4 s^2 e^{-2 s^2 \left(k_1^2+k_2^2\right)}}{\pi } \;\;\;
\begin{blockarray}{ccccc}
 LL & LR & RL & RR & \\ 
\begin{block}{(cccc)c}
 \frac{1}{4} & 0 & 0 &~-\frac{1}{4} e^{-2 i d (k_1+k_2)} & ~~LL \\
 & 0 & 0 & 0 & ~~LR \\
  \BAmulticolumn{2}{c}{\multirow{2}{*}{{\LARGE h.c.}}} & 0 & 0 & ~~RL \\
 & & & \frac{1}{4} & ~~RR \\
\end{block}
\end{blockarray}
\end{align}

\newpage

\subsection{3rd excited state}
\label{a46}

\noindent $\rho_{\rm H}^{S=0,S_z=0}$ and $\eta_{\rm Hub}^{S=0,S_z=0}(k_1,k_2)$ for the third excited state 
of the Hubbard Hamiltonian [see Eq.\ (\ref{eq:hubbard3rdvariant2})] are given by:

\begin{align}
\arraycolsep=1.4pt\def\arraystretch{2.2}
\rho_{\rm Hub}^{S=0,S_z=0}&=\frac{1}{16-\mathcal{P}(\mathcal{U})\mathcal{U}} \;\;\;
\begin{blockarray}{ccccc}
 LL & LR & RL & RR & \\ 
\begin{block}{(cccc)c}
 4 & -\mathcal{P}(\mathcal{U}) & -\mathcal{P}(\mathcal{U}) & 4 & ~~LL \\
 & ~~4-\frac{\mathcal{P}(\mathcal{U})\mathcal{U}}{2}~~ & ~~4-\frac{\mathcal{P}(\mathcal{U})\mathcal{U}}{2}~~ & -\mathcal{P}(\mathcal{U}) & ~~LR \\
  \BAmulticolumn{2}{c}{\multirow{2}{*}{{\LARGE h.c.}}} & ~~4-\frac{\mathcal{P}(\mathcal{U})\mathcal{U}}{2}~~ & -\mathcal{P}(\mathcal{U}) & ~~RL \\
 & & & 4 & ~~RR \\
\end{block}
\end{blockarray},\\
&=\;\;\begin{blockarray}{ccccc}
 LL & LR & RL & RR & \\ 
\begin{block}{(cccc)c}
 C(\mathcal{U})^2 & -C(\mathcal{U})D(\mathcal{U}) & -C(\mathcal{U})D(\mathcal{U}) & C(\mathcal{U})^2 & ~~LL \\
 & D(\mathcal{U})^2 & D(\mathcal{U})^2 &  -C(\mathcal{U})D(\mathcal{U}) & ~~LR \\
 \BAmulticolumn{2}{c}{\multirow{2}{*}{{\LARGE h.c.}}} & D(\mathcal{U})^2 &  -C(\mathcal{U})D(\mathcal{U}) & ~~RL \\
 & & & C(\mathcal{U})^2 & ~~RR \\
\end{block}
\end{blockarray}.
\end{align}
 
where $\mathcal{U}=U/t$ and  $\mathcal{P}(\mathcal{U})=\sqrt{16+\mathcal{U}^2}-\mathcal{U}$.
 
{\footnotesize
\begin{align}
& \eta_{\rm Hub}^{S=0,S_z=0}(k_1,k_2)= \nonumber \\
\arraycolsep=1.4pt\def\arraystretch{1.8}
& \frac{2 s ^2 e^{-2 s ^2 \left(k_1^2+k_2^2\right)}}{\pi  (\mathcal{P}(\mathcal{U}) U-16)} \;\;\;
\begin{blockarray}{ccccc}
 LL & LR & RL & RR & \\ 
\begin{block}{(cccc)c}
 4 & ~-e^{-2 i d k_2} \mathcal{P}(\mathcal{U})~ & -e^{-2 i d k_1} \mathcal{P}(\mathcal{U}) & 4 e^{-2 i d (k_1+k_2)} & ~~LL \\
 & 4-\frac{\mathcal{P}(\mathcal{U})\mathcal{U}}{2} & ~-\frac{1}{2} e^{-2 i d (k_1-k_2)} (\mathcal{P}(\mathcal{U}) U-8)~~ & -e^{-2 i d k_1} \mathcal{P}(\mathcal{U}) & ~~LR \\
  \BAmulticolumn{2}{c}{\multirow{2}{*}{{\LARGE h.c.}}} & 4-\frac{\mathcal{P}(\mathcal{U})\mathcal{U}}{2} & -e^{-2 i d k_2} \mathcal{P}(\mathcal{U}) & ~~RL \\
 & & & 4 & ~~RR \\
\end{block}
\end{blockarray}.
\end{align}
}

\end{widetext}


\begin{thebibliography}{99}
\bibitem{bloc05}
S. F\"{o}lling, F. Gerbier, A. Widera, O. Mandel, T. Gericke, and I. Bloch,
Spatial quantum noise interferometry in expanding ultracold atom clouds,
Nature {\bf 434}, 481 (2005). 
\bibitem{bloc06.2}
T. Rom, Th. Best, D. van Oosten, U. Schneider, S. F\"{o}lling, B. Paredes, and I. Bloch,
Free fermion antibunching in a degenerate atomic Fermi gas released from an optical lattice,
Nature {\bf 444}, 733 (2006). 
\textcolor{black}{
\bibitem{kauf14}
A.M. Kaufman, B.J. Lester, C.M. Reynolds, M.L. Wall, M. Foss-Feig, 
K.R.A. Hazzard, A.M. Rey, and C.A. Regal,
Two-particle quantum interference in tunnel-coupled optical tweezers,
Science {\bf 345}, 306 (2014).
}
\bibitem{bouc16}
B. Fang, A. Johnson, T. Roscilde, and I. Bouchoule,
Momentum-Space Correlations of a One-Dimensional Bose Gas,
Phys. Rev. Lett. {\bf 116}, 050402 (2016).
\bibitem{hodg17}
S.S. Hodgman, R.I. Khakimov, R.J. Lewis-Swan, A.G. Truscott, and K.V. Kheruntsyan,
Solving the Quantum Many-Body Problem via Correlations Measured with a Momentum Microscope,
Phys. Rev. Lett. {\bf 118}, 240402 (2017).
\bibitem{schm17}
M. Bonneau, W.J. Munro, K. Nemoto, and J. Schmiedmayer,
Characterizing two-particle entanglement in a double-well potential,
arXiv:1711.08977.
\bibitem{berg17}
A. Bergschneider,
Strong correlations in few-fermion systems,
(Ph.D. Dissertation, Ruperto-Carola University of Heidelberg, 2017), 
DOI: 10.11588/heidok.00023328.
\bibitem{bloc06}
P. Treutlein, T. Steinmetz, Y. Colombe, B. Lev, P. Hommelhoff, J. Reichel, M. Greiner, O. Mandel,
A. Widera, T. Rom, I. Bloch, and Th.W. H\"{a}nsch,
Quantum information processing in optical lattices and magnetic microtraps,
Fortschr. Phys. {\bf 54}, 702 (2006). 
\textcolor{black}{
\bibitem{isla15}
R. Islam, R. Ma, Ph.M. Preiss, M.E. Tai, A. Lukin, M. Rispoli, and M. Greiner,
Measuring entanglement entropy in a quantum many-body system,
Nature {\bf 528}, 77 (2015).
}
\bibitem{bloc04}
B. Paredes, A. Widera, V. Murg, O. Mandel, S. F\"{o}lling, I. Cirac, G.V. Shlyapnikov,
Th.W. H\"{a}nsch, and I. Bloch, 
Tonks-Girardeau gas of ultracold atoms in an optical lattice, 
Nature {\bf 429}, 277 (2004).
\bibitem{yann17}
B.B. Brandt, C. Yannouleas, and U. Landman,
Two-point momentum correlations of few ultracold quasi-one-dimensional trapped
fermions: Diffraction patterns,
Phys. Rev. A {\bf 96}, 053632 (2017).
\bibitem{joch12}
G. Z\"{u}rn, F. Serwane, T. Lompe, A.N. Wenz, M.G. Ries, J.E. Bohn, and S. Jochim,
Fermionization of two distinguishable fermions,
Phys. Rev. Lett. {\bf 108}, 075303 (2012).
\bibitem{joch15}
S. Murmann, A. Bergschneider, V.M. Klinkhamer, G. Z\"{u}rn, T. Lompe, and S. Jochim,
Two fermions in a double well: Exploring a fundamental building block of the Hubbard model,
Phys. Rev. Lett. {\bf 114}, 080402 (2015).
\bibitem{yann15}
B.B. Brandt, C. Yannouleas, and U. Landman,
Double-well ultracold-fermions computational microscopy: Wave-function anatomy of attractive-pairing 
and Wigner-molecule entanglement and natural orbitals,
Nano Lett. {\bf 15}, 7105 (2015).
\textcolor{black}{
\bibitem{hong87}
C. K. Hong, Z. Y. Ou, and L. Mandel,
Measurement of subpicosecond time intervals between two photons by interference,
Phys. Rev. Lett. {\bf 59}, 2044 (1987).
\bibitem{ou07}
Z. Y. Ou, 
Multi-photon interference and temporal distinguishability of photons,
Int. J. Mod. Phys. {\bf 21}, 5033 (2007).
\bibitem{liu98}
R.C. Liu, B. Odom, Y. Yamamoto, and S. Tarucha,
Quantum interference in electron collision,
Nature {\bf 391}, 263 (1998).
\bibitem{bocq13}
E. Bocquillon, V. Freulon, J.-M. Berroir, P. Degiovanni, B. Pla\c{c}ais, 
A. Cavanna, Y. Jin, and G. F\`{e}ve,
Coherence and indistinguishability of single electrons emitted by independent sources,
Science {\bf 339}, 1054 (2013).
\bibitem{aspe15}
R. Lopes, A. Imanaliev, A. Aspect, M. Cheneau, D. Boiron, and C.I. Westbrook,
Atomic Hong-Ou-Mandel experiment, 
Nature {\bf 520}, 66 (2015).
}
\bibitem{yann16}
C. Yannouleas, B.B. Brandt, and U. Landman,
Ultracold few fermionic atoms in needle-shaped double wells: Spin chains and resonating spin clusters
from microscopic hamiltonians emulated via antiferromagnetic Heisemberg and $t$-$J$ models,
New J. Phys. {\bf 18}, 073018 (2016). 
\bibitem{shul56}
P.-O. L\"{o}wdin and H. Shull,
Natural orbitals in the quantum theory of two-electron systems,
Phys. Rev. {\bf 101}, 1730 (1956).
\bibitem{aspe07}
T. Jeltes, J.M. McNamara, W. Hogervorst, W. Vassen, V. Krachmalnicoff, M. Schellekens, A. Perrin,
H. Chang, D. Boiron, A. Aspect, and C.I. Westbrook,
Comparison of the Hanbury Brown-Twiss effect for bosons and fermions,
Nature {\bf 445}, 402 (2007).
\bibitem{ou88}
Z.Y. Ou and L. Mandel,
Observation of spatial quantum beating with separated photodetectors,
Phys. Rev. Lett. {\bf 61}, 54 (1988).
\bibitem{altm04}
E. Altman, E. Demler, and M.D. Lukin,
Probing many-body states of ultracold atoms via noise correlations,
Phys. Rev. A {\bf 70}, 013603 (2004).
\bibitem{blan00}
Ya.M. Blanter and M. B\"{u}ttiker,
Shot noise in mesoscopic conductors,
Physics Reports {\bf 336}, 1 (2000).
\bibitem{note1}
The preparatory measurements were performed for three different interaction strengths $U=0$ 
(non-interacting), $U/t=2.1$ (weak strength), and $U/t=7.7$ (intermediate strength); see Fig.\ 7.8(a-c) in 
Ref.\ \cite{berg17}. In the absence of the present CI results and the Hubbard-type formula
in Eq.\ (\ref{hubgs}), no analysis of the full correlation maps and their interference patterns was carried
out in Ref.\ \cite{berg17}. 
\bibitem{sneg}
R. \v{Z}itko,
SNEG $-$ Mathematica package for symbolic calculations with second-quantization-operator expressions,
Comput. Phys. Commun. {\bf 182}, 2259 (2011).
\bibitem{math}
MATHEMATICA
(Wolfram Research, Inc., Version 11.2, Champaign, IL, 2017).
\bibitem{yann07}
C. Yannouleas and U. Landman,
Symmetry breaking and quantum correlations in finite systems: Studies of quantum dots and
ultracold bose gases and related nuclear and chemical methods,
Rep. Prog. Phys. {\bf 70}, 2067 (2007).
\bibitem{li09}
Ying Li, C. Yannouleas, and U. Landman,
Artificial quantum-dot helium molecules: Electronic spectra, spin structures, and Heisenberg clusters,
Phys. Rev. B {\bf 80}, 045326 (2009).
\bibitem{schl01}
J. Schliemann, D. Loss, and A.H. MacDonald,
Double-occupancy errors, adiabaticity, and entanglement of spin qubits in quantum dots,
Phys. Rev. B {\bf 63}, 085311 (2001).
\bibitem{schl02}
K. Eckert, J. Schliemann, D. Bru{\ss}, and M. Lewenstein,
Quantum correlations in systems of indistinguishable particles,
Annals of Physics (New York) {\bf 299}, 88 (2002).
\end{thebibliography}
\end{document}